\documentclass[aps,pra,amsmath,twocolumn,amssymb,footinbib,superscriptaddress]{revtex4}

\usepackage{ulem}
\usepackage{graphicx,amsfonts}
\usepackage{dcolumn}
\usepackage{bm}

\newcommand{\ket}[1]{\displaystyle{|#1\rangle}}
\newcommand{\bra}[1]{\displaystyle{\langle #1|}}
\newcommand{\proj}[2]{\displaystyle{| #1 \rangle \langle #2|}}

\usepackage{pstricks}

\begin{document}
\normalem
\preprint{APS/123-QED}

\title{Emerging Bosons with Three-Body Interactions from Spin-1 Atoms in Optical Lattices}
\date{\today}
\author{L. Mazza}
\affiliation{Max-Planck-Institut f\"ur Quantenoptik, Hans-Kopfermann-Stra\ss e 1, D-85748 Garching, Germany}
\email{leonardo.mazza@mpq.mpg.de}
\author{M. Rizzi}
\affiliation{Max-Planck-Institut f\"ur Quantenoptik, Hans-Kopfermann-Stra\ss e 1, D-85748 Garching, Germany}
\author{M. Lewenstein}
\affiliation{ICFO - Institut de Ci\`encies Fot\`oniques, Parc Mediterrani de la Tecnologia, E-08860 Castelldefels (Barcelona), Spain}
\affiliation{ICREA - Instituci\'o Catalana de Ricerca i Estudis Avan\c{c}ats, E-08010 Barcelona, Spain}
\author{J.I. Cirac}
\affiliation{Max-Planck-Institut f\"ur Quantenoptik, Hans-Kopfermann-Stra\ss e 1, D-85748 Garching, Germany}

\date{\today}

\begin{abstract}

We study two many-body systems of bosons interacting via an infinite three-body contact repulsion in a lattice: 
a pairs quasi-condensate induced by correlated hopping and
the discrete version of the Pfaffian wavefunction.
We propose to experimentally realise systems characterized by such interaction by means of a proper spin-1 lattice Hamiltonian: spin degrees of freedom are locally mapped into occupation numbers of emerging bosons,
in a fashion similar to spin-1/2 and hardcore bosons. Such a system can be realized with ultracold spin-1 atoms 
in a Mott Insulator with filling-factor one. The high versatility of these setups allows us to engineer  spin-hopping operators breaking
the SU(2) symmetry, as needed to approximate interesting bosonic Hamiltonians with three-body hardcore constraint. For this purpose
we combine bichromatic spin-independent superlattices and Raman transitions to induce a different hopping rate for each spin orientation.
Finally, we illustrate how our setup could be used to experimentally realize the first setup, i.e. the transition to a pairs quasi-condensed
phase of the emerging bosons. We also report on a route towards the realization of a discrete bosonic Pfaffian wavefunction
and list some open problems to reach this goal.

\end{abstract}

\maketitle

\section{Introduction}

Ultracold atoms and trapped ions offer unprecedented possibilities to realize, control and observe quantum
many-body phenomena~\cite{Jaksch05, Bloch08, Lewenstein07}. For these reasons, in the last years they have been frequently employed
as quantum simulators, i.e. controllable laboratory setups mimicking other interesting but not easily accessible
systems described by the same mathematical model. Simulation targets come from diverse research fields, as condensed-matter or even
high-energy physics. 
A paradigmatic quantum simulator is a system of bosonic atoms in an optical
lattice, which provides a practically ideal realization of the Bose-Hubbard 
model~\cite{Fisher89, Jaksch98, Greiner02, Trotzky09}. On one hand, many of these studies aim at the simulation of systems which
are otherwise difficult to treat numerically, such as
the Fermi-Hubbard model~\cite{Jordens08, Schneider08} or many-body frustrated models~\cite{Lewenstein07}.
On the other hand, a parallel goal of quantum simulation is the experimental study of ``blackboard''
theoretical models, such as the Ising model~\cite{Friedenauer08},
the Tonks-Girardeau gas~\cite{Paredes04}, or the one-dimensional Dirac equation~\cite{Blatt2010}.

Within the last context, the possibility of simulating many-body systems characterized by 
interactions involving more than two particles is of great interest. For example, three- and four-body
 interactions are known to be the essential ingredients of lattice gauge theories~\cite{Montvay97}.
Such theories and their related lattice models play important roles
in the context of novel exotic quantum phases, of the breakdown of the Landau-Ginzburg scenario
and of the confinement-deconfinement transition~\cite{sachdevNPhys, Alet2006122, PhysRevLett.98.227202}.
Besides that, local many-body interactions are essential also in paradigmatic spin-models exhibiting topological
order~\cite{Kitaev20062,PhysRevB.75.075103,PhysRevB.71.045110}.
Moreover, in the presence of external magnetic fields,  they
lead to various exotic fractional quantum Hall states~\cite{Ezawa00, cooper}.
A celebrated example is the 
Pfaffian wavefunction~\cite{pfaffian, Greiter1992567},
which arises in bosonic systems at filling factor $\nu=1$ and exhibits topological order.
Its quasi-excitations are non-Abelian anyons, i.e. the exchange of these quasiparticles is associated with
non-commuting trasformation of the system~\cite{Jacak03}.
Furthermore, Pfaffian states 
can be found within other frameworks,
such as p-wave superconductivity, where the excitations correspond to zero-energy 
Majorana fermions~\cite{pfaffian, RevModPhys.80.1083, schrieffer}, or one-dimensional systems~\cite{PhysRevA.75.053611}.
This motivates the need for experimentally feasible proposals realising three- or many-body interactions.

Recently, cold atom theorists have been developing several approaches to achieve this goal.
The early proposals employed higher-order super-exchange interactions on triangular
and kagom\'e lattices~\cite{PhysRevA.70.053620}.
Alas, the temperatures required are even lower than those necessary for quantum magnetism in Mott phases, 
a demanding task to which a lot of experimentalists are still committed.
Super-exchange interactions of the second order involving Raman transitions 
between atoms and molecules in square lattices have been proposed to realize
an effective ring-exchange Hamiltonian for bosons~\cite{Buechler05}. A completely different approach has been proposed by B\"uchler et al.~\cite{Buchler07b} who suggested to use polar molecules with a setup inhibiting two-body interactions.
Very recently it has been suggested to use the dissipative dynamics in presence of a large three-body loss rate 
in order to enforce on the system an effective three-body hardcore constraint~\cite{daley:040402, roncaglia}.
The idea is very reminiscent of a scheme experimentally realized by Syassen et al.~\cite{Syassen09} 
to induce strong correlations in molecular gases. The dissipative scheme has been combined with the rotation of the trap
to induce artificial magnetic fields and achieve the formation of the Pfaffian wavefunction~\cite{roncaglia}.
Finally, it has been theoretically shown that a perturbative treatment of vibrational bands in optical lattices induces effective many-body terms in the Hamiltonian, which have also been experimentally observed in the time-evolution of the quantum phase of the system~\cite{1367-2630-11-9-093022, 3bodybloch}.

The purpose of this paper is twofold. The first one is the investigation of two 
systems of bosons interacting via infinite three-body repulsion in a lattice. 
We show that in one dimension the presence of correlated hopping in such models leads to an exotic quantum phase, 
characterised by quasi-condensation of pairs of particles. In the presence of magnetic fields in 2D,
the ground state of such model corresponds to the lattice version of the bosonic Pfaffian wavefunction.
In particular, we discuss the stability of the topological properties of such wavefunction
in a non-dilute limit with the magnetic length comparable to the lattice spacing.

The second purpose is to illustrate how spin-1 atoms in a Mott Insulator (MI) with filling one offer a possibility to 
tailor three-body interactions in a setup combinable with artificial gauge fields.
The idea is based on mapping the internal states of real spin-1 bosons on the lattice into occupation numbers
of emerging bosons, similarly to the correspondence between spin-1/2 particles and emerging hardcore bosons.
An easy generalization to higher-spin atoms can open the route to the simulation of four-, five- and many-body contact infinite repulsions.
The first proposed model, which exhibits the quasi-condensation of pairs, can be carried on rather simply in such a  framework. On the other side, the realisation of the lattice Pfaffian wavefunction seems to be slightly outside the class of models accessible with our proposal.

\begin{figure}[t]
\includegraphics[width=8.0cm]{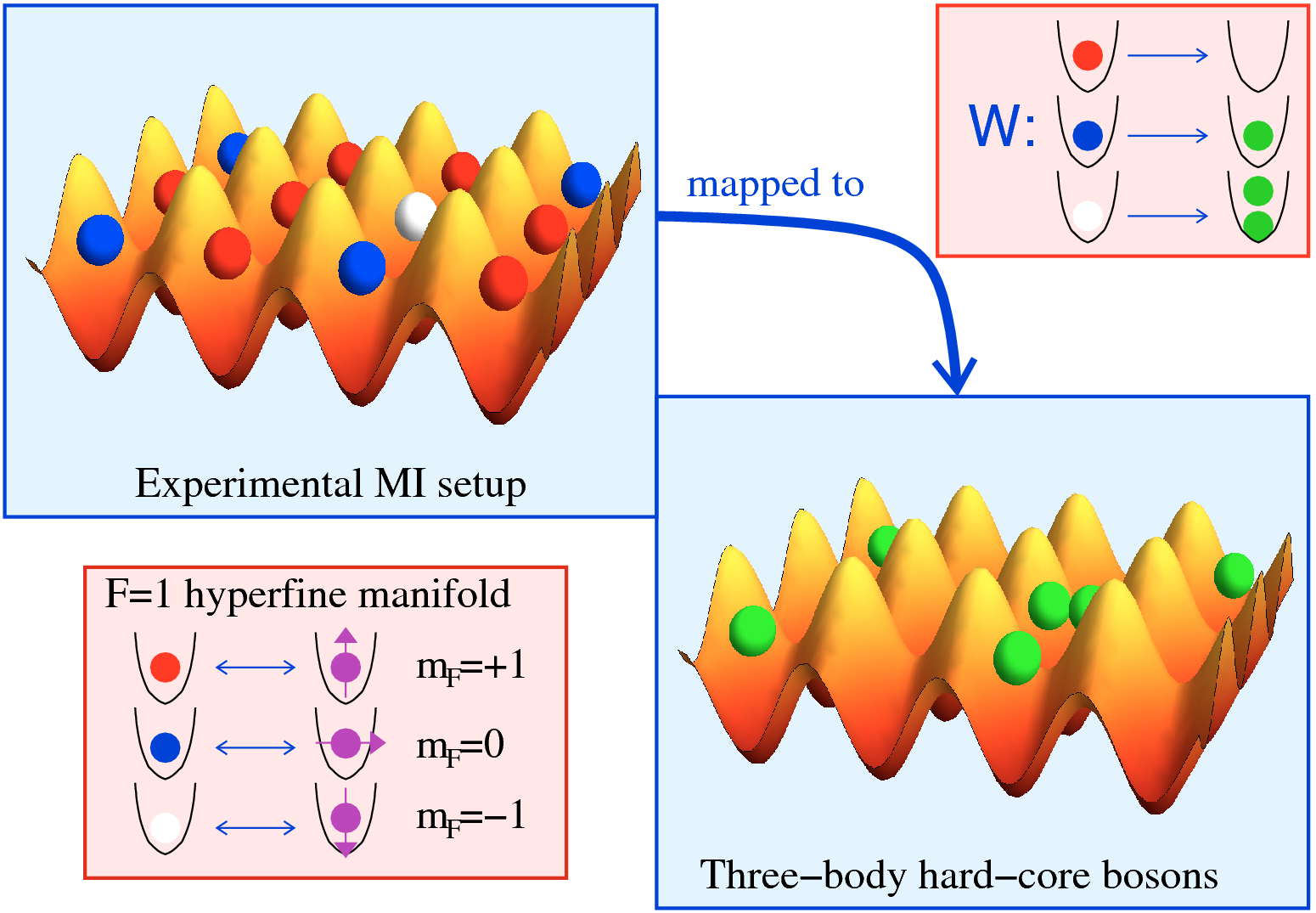}
\caption{Sketch of the proposed mapping. A Mott Insulator with filling $1$ whose atoms have three relevant degrees of freedom
can simulate a system of bosons with an infinite contact repulsion via the mapping $\mathsf W$.
The three relevant degrees of freedom can be identified with a $F=1$ hyperfine manifold.}
\label{fig:mapping}
\end{figure}

The experimental system discussed in this article greatly exploits the use of bichromatic superlattices~\cite{superlatticefolling} 
in order to individually tailor the hopping rates of the different spin species.
This scheme is rather general and could allow the realization of very general hopping operators, including even spin-flipping terms;
therefore, it could find applications beyond the problem of many-body interactions, such as in fermionic systems
or in non-Abelian gauge theories~\cite{bermudez}.

The article is organised as follows: 
we start in Sec.~\ref{sec:mapping} describing how to realise three-body interacting bosons
with atomic setups characterized by three relevant degrees of freedom.
Subsequently, in Sec.~\ref{sec:spin1atoms}, we consider the explicit case of atoms with $F=1$ hyperfine ground manifold
and we derive the related super-exchange Hamiltonian.
We then move to the problem of having a complete external access to all the relevant parameters
of the real spin-1 system and in Sec.~\ref{sec:experimental} we describe how superlattices and Raman transitions can be
used to achieve this goal. 
The next two sections are devoted to the discussion of systems characterized by three-body interactions and they possible implementation with our proposal.
In Sec.~\ref{sec:pqc} we present  a phase characterized by quasi-long-range order of pairs.
In Sec.~\ref{sec:pfaffian} we discuss the problems faced in trying to engineer the Pfaffian
wavefunction. Finally, in Sec.~\ref{sec:conclusions} our conclusions are presented.

\section{The mapping} \label{sec:mapping}

As a starting point for our work, we recall here that particles interacting via a three-body infinite repulsion
effectively undergo a three-body hardcore constraint, i.e. there can not be more than two particles at a time in the same place.
Therefore, in the presence of a spatially discrete setup, the description of local (on-site) degrees of freedom
is captured by a finite Hilbert space of dimension three:
\begin{equation}
\mathcal{H}^{\mathrm{loc}}_{\mathrm{3hb}} = \mathrm{Span} \{ \ket{n=0}, \ket{n=1}, \ket{n=2} \}
\end{equation}

Since three-body elastic interactions are rather weak in nature, we have to simulate such an ideal system by means of 
a proper realistic discrete setup characterized by a local Hilbert space $\mathcal{H}^{\mathrm{loc}}_{\mathrm{real}}$ of dimension three. 
A unitary mapping between the local Hilbert spaces (Fig.\ref{fig:mapping})
\begin{equation}
\mathsf W: \mathcal{H}^{\mathrm{loc}}_{\mathrm{real}} \longrightarrow  \mathcal{H}^{\mathrm{loc}}_{\mathrm{3hb}}, 
\label{eq:generalmap}
\end{equation}
allows us to relate the real experimental dynamics to the dynamics of the emerging bosons characterized
by an infinite three-body interaction.
If the first is described by some effective Hamiltonian $H^{\mathrm{eff}}_{\mathrm{real}}$, the correspondent $H_{\mathrm{3hb}}$ would then be
\begin{equation}
H_{\mathrm{3hb}} = \mathsf W^{\otimes L^2} \; H^{\mathrm{eff}}_{\mathrm{real}} \; \, \mathsf{W}^{\dagger \, \otimes L^2} \; ,
\end{equation}
and tuning experimental parameters in $H^{\mathrm{eff}}_{\mathrm{real}}$ permits in principle
the investigation of a variety of ``blackboard'' $H_{\mathrm{3hb}}$.

In the context of ultracold atoms in optical lattices, which offer an unprecedented control on discrete structures, it is then quite natural
to consider a Mott Insulator (MI) with filling $1$ and three internal degrees of freedom. 
In this article we focus our attention on an hyperfine manifold $F=1$, as exhibited by $^{87}$Rb or $^{23}$Na,
that have already been successfully loaded into an optical lattice and cooled into a MI state without freezing
the spin-dynamics~\cite{1367-2630-8-8-152}. 
If the experimental apparatus is tuned to a not too deep MI, small quantum fluctuations in the atomic position
give rise to a non-trivial dynamics, usually addressed as superexchange effects. This dynamics can be characterized
by an effective Hamiltonian $H^{\mathrm{eff}}_{\mathrm{real}}$ obtained through a second-order perturbative expansion
of the kinetic term of the Bose-Hubbard Hamiltonian describing the real atoms on the lattice \cite{PhysRevLett.81.3108}.
The next Section is indeed devoted to a detailed derivation of such effective theory.
However, we underline that this is not the only possible choice and other internal degrees of freedom could have been chosen;
as proposed in Ref.~\cite{Svistunov} for a different purpose, these three local degrees of freedom could even
correspond to different scalar atomic species.

\section{Spin 1 Atoms} \label{sec:spin1atoms}

The focus of this Section consists in deriving the effective Hamiltonian describing magnetic degrees of freedom 
inside the Mott Insulator and in investigating whether interesting $H_{\mathrm{3hb}}$ Hamiltonians
can be effectively mimicked via the proposed mapping.
The proposed analysis is generally valid for systems of dimension one, two and three.
In order to describe the experimental system of ultracold spin-1 atoms in an optical lattice, we rely on the standard
Bose-Hubbard Hamiltonian~\cite{PhysRevA.68.063602} ($\alpha=\{-,\circ,+\}$ runs over the three spin states
$\{ \ket{m_F=-1}, \ket{m_F=0}, \ket{m_F=+1} \}$) :
\begin{eqnarray}
H_{\mathrm{real}} &=&  \sum_{<i,j>, \alpha} [ - t_{\alpha} b^{\dagger}_{i,\alpha} b_{j,\alpha}+ H.c. ] 
+ \sum_{i, \alpha} \Delta_{\alpha} n_{i, \alpha} + \nonumber \\ 
&+& \frac{U_0}{2} \sum_i n_i (n_i - 1) + \frac{U_2}{2} \sum_i (\vec{S}_i^2 - 2 n_i)
\label{eq:ham:spin1:clean}
\end{eqnarray}
with $n_{i,\alpha} = b^{\dagger}_{i, \alpha} b_{i, \alpha}$, $n_i = \sum_{\alpha} n_{i,\alpha}$
and $( \vec{S}_i )_{\alpha,\beta} = b^{\dagger}_{i, \alpha} \vec{F}_{\alpha,\beta}b_{i, \beta}$ is the total spin on the site $i$.
The first term represents the kinetic energy (hopping) and the $\Delta_{\alpha}$'s represent the energy offset
of each of the three states ($\Delta_{\circ} = 0$);
the last two terms describe the two-body interaction on a same site that, due to the spin nature of atoms,
is characterized by two s-wave scattering lengths, $a_0$ for the $S^{\mathrm{tot}}=0$ channel and $a_2$ for the $S^{\mathrm{tot}}=2$ one,
with the ratio $U_2/U_0 = (a_2 - a_0)/(a_2 + 2 a_0)$~\cite{PhysRevA.68.063602}.

Hamiltonian (\ref{eq:ham:spin1:clean}) preserves the total magnetization $M \equiv \sum_i S^{z}_{i} = \sum_i \alpha_i$ of the sample,
allowing us to work in a convenient block-diagonal representation.
Moreover, each energy offset $\Delta_{\alpha}$ plays the role of a chemical potential for the atomic species $\alpha$.
As a consequence, in absence of spin-flipping interactions ($n_{\alpha}$ conserved), the  $\Delta_{\alpha}$ would play no role in the dynamics at fixed magnetization. 
In our case the situation is complicated by the presence, in the atomic Hamiltonian, of terms which flip the atomic spin:
\begin{equation}
\ket{m_F = 0} \ket{0} \; \longleftrightarrow \; \ket{+1} \ket{-1}
\end{equation}
The $\Delta_{\alpha}$ would still not play any role in the dynamics if the following relation holds:
$2 \Delta_{\circ} = \Delta_{+} + \Delta_{-}$.
In presence of an external magnetic field, only the linear Zeeman shift satisfies this requirement, whereas the quadratic one does not. In this case, the relevant dynamical quantity is $\delta = \Delta_{+} + \Delta_{-} - 2 \Delta_{\circ}$, which quantifies deviations from the linear splitting regime. It is experimentally possible to control small values of $\delta$ dressing the system with microwave fields~\cite{PhysRevA.73.041602}.

In case interaction energies are larger than the hopping rates ($U_0 + U_2,U_0 - 2 U_2 \gg |t_{\alpha}|$),
the system is in a MI phase and with a second-order perturbative expansion of the kinetic term we compute
the super-exchange Hamiltonian $H^{\mathrm{eff}}_{\mathrm{real}}$, whose link-expression reads as follows:

\begin{widetext}
\begin{displaymath}
\text{Link basis \textit{real}: } \{ \ket{m_{F,\mathrm{site 1}} = -} \ket{m_{F,\mathrm{site 2}} = -}, 
\, \ket{-}\ket{\circ}, \, \ket{-}\ket{+},
\, \ket{\circ}\ket{-},  \, \ket{\circ}\ket{\circ}, \, \ket{\circ}\ket{+},
\ket{+}\ket{-}, \, \ket{+}\ket{\circ},
\ket{+}\ket{+}   \}
\end{displaymath}
\begin{displaymath}
H^{\mathrm{eff}}_{\mathrm{real}} = 
H^{\mathrm{eff} (0)}_{\mathrm{real}}
+
H^{\mathrm{eff} (2)}_{\mathrm{real}} \qquad \qquad
H^{\mathrm{eff} (0)}_{\mathrm{real}} = \mathrm{Diag}
\left(
2 \Delta_{-}, \, \Delta_{-}, \, \Delta_{+} + \Delta_{-}, \, \Delta_{-}, \, 0, \, \Delta_{+}, \, \Delta_{+} + \Delta_{-}, 
\, \Delta_{+}, \, 2 \Delta_{+}
\right) 
\end{displaymath}
\begin{equation}
H^{\mathrm{eff} (2)}_{\mathrm{real}}  = - \frac{1}{U_0 + U_2} \left(
\begin{array}{ccccccccc}
\squeezetable
{\scriptstyle 4 |t_{-}|^2}  & \hspace{-0.2cm} 0 & \hspace{-0.3cm} 0 & \hspace{-0.2cm} 0 & \hspace{-0.4cm} 0 &  0 & 0 & \hspace{-0.3cm} 0 & \hspace{-0.2cm} 0  \\
0 & \hspace{-0.2cm} {\scriptstyle |t_{-}|^2 + |t_{\circ}|^2} & \hspace{-0.3cm} 0 & \hspace{-0.2cm} {\scriptstyle 2 t_{\circ}^* t_{-} } & \hspace{-0.4cm} 0 & 0 & 0 & \hspace{-0.3cm} 0 & \hspace{-0.2cm} 0  \\
0 & \hspace{-0.2cm} 0 & \hspace{-0.3cm} {\scriptstyle (|t_{-}|^2 + |t_{+}|^2) \mathcal B } & \hspace{-0.2cm} 0 & \hspace{-0.4cm}
{\scriptstyle-(t_{\circ}^* t_{-} + t_+^* t_{\circ})  \mathcal A } & 0 & 
{\scriptstyle 2 t_+^* t_{-}  \mathcal B }  & \hspace{-0.3cm} 0 & \hspace{-0.2cm} 0 \\
0 & \hspace{-0.2cm} {\scriptstyle 2 t^*_{-} t_{\circ} } & \hspace{-0.3cm} 0 & \hspace{-0.2cm} {\scriptstyle |t_{-}|^2 + |t_{\circ}|^2 }& \hspace{-0.4cm} 0 & 0 & 0 & \hspace{-0.3cm} 0 & \hspace{-0.2cm} 0 \\
0 & \hspace{-0.2cm} 0 & \hspace{-0.3cm} {\scriptstyle - (t_{\circ}^* t_+ + t^*_{-} t_{\circ})  \mathcal A }& \hspace{-0.2cm} 0 & \hspace{-0.4cm}
  {\scriptstyle |t_{\circ}|^2 \mathcal C}  & 0 & 
{\scriptstyle -(t_{\circ}^* t_{-}  + t_+^* t_{\circ})  \mathcal A } & \hspace{-0.3cm} 0 & \hspace{-0.2cm} 0 \\
0 & \hspace{-0.2cm} 0 & \hspace{-0.3cm} 0 & \hspace{-0.2cm} 0 & 0 & \hspace{-0.4cm} {\scriptstyle |t_{\circ}|^2 + |t_{1}|^2 } & 0 & \hspace{-0.3cm} {\scriptstyle 2 t_{+}^* t_{\circ} } & \hspace{-0.2cm} 0\\
0 & \hspace{-0.2cm} 0 & \hspace{-0.3cm} {\scriptstyle 2 t_{-}^* t_+  \mathcal B} & \hspace{-0.2cm} 0 & \hspace{-0.4cm}
{\scriptstyle - (t_{-}^* t_{\circ}  + t_{\circ}^* t_+)  \mathcal A } & 0  &
{\scriptstyle (|t_+|^2 + |t_{-}|^2 ) \mathcal B } & \hspace{-0.3cm} 0 & \hspace{-0.2cm} 0 \\
0 & \hspace{-0.2cm} 0 & \hspace{-0.3cm} 0 & \hspace{-0.2cm} 0 & \hspace{-0.4cm} 0 & {\scriptstyle 2 t_{\circ}^* t_{+} } &0&\hspace{-0.3cm} {\scriptstyle |t_{\circ}|^2 + |t_{+}|^2 } & \hspace{-0.2cm}0 \\
0 & \hspace{-0.2cm} 0 & \hspace{-0.3cm} 0 & \hspace{-0.2cm} 0 & \hspace{-0.4cm} 0 & 0 & 0 & \hspace{-0.3cm} 0 & \hspace{-0.2cm} {\scriptstyle 4 |t_{+}|^2 }
\end{array}
\right) 
\label{eq:bigmatrix}
\end{equation}
\begin{equation}
{\textstyle 
\mathcal A = 
\frac{U_2 (U_0 + U_2)}{(U_0 + U_2)(U_0 - 2 U_2) + \delta U_0} + \frac{U_2 (U_0 + U_2)}{(U_0 + U_2)(U_0 - 2 U_2) + \delta (U_2 - U_0)} 
\qquad
\mathcal B = \frac{(U_0 - \delta) (U_0 + U_2)}{(U_0 + U_2)(U_0 - 2 U_2) + \delta (U_2 - U_0)}
\qquad
\mathcal C = \frac{4(U_0 - U_2 + \delta)(U_0 + U_2)}{(U_0 + U_2) (U_0 - 2 U_2 ) + \delta U_0}
} \nonumber
\end{equation}
\end{widetext}

Let us now consider a simple class of mappings $W_{\{\varphi\}}$  which will be used in the following, characterized only by a simple phase freedom:
\begin{equation}
\mathsf W_{\{\varphi\}}  \ket{m_F = \alpha}_{\mathrm{real}}  = e^{i \varphi_{\alpha}} \ket{n = \alpha + 1}_{\mathrm{3hb}}
\label{eq:mapping}
\end{equation}
These mappings are characterized by the property that the  magnetization of the spin insulator is directly mapped into the density of the three-hardcore bosons. Therefore, since the Hamiltonian in Eq.~(\ref{eq:bigmatrix}) contains only off-diagonal terms which preserve the total magnetization, we automatically gain the possibility of studying hardcore bosons setups at fixed density.

As far as the interaction strengths are concerned, we report that the scattering lengths $a_0$ and $a_2$ have very similar values both in $^{87}$Rb and $^{23}$Na. This means that the spin-dependent part of the interaction is in natural setups almost negligible, as it is clearly stated once the ratio $U_2/U_0$ is calculated, respectively $-0.005$ and $0.04$ for the two atoms~\cite{PhysRevLett.88.093201, 1367-2630-8-8-152, phillips}. This could in principle be a weak point of our proposal, because  four non-diagonal matrix element in $H^{\mathrm{eff} }_{\mathrm{ real}}$ are proportional to $U_2$. In the next sections we will show how to cope with this problem, and namely that even in the case in which it is not possible to externally tune the ratio $U_2/U_0$ to higher values it is possible to experimentally observe some interesting physics.

Before concluding this section, we would like to stress that the off-diagonal matrix elements present in Eq.~(\ref{eq:bigmatrix})  correspond, via the mappings  $W_{\{\varphi\}}$  in Eq.~(\ref{eq:mapping}), to terms describing the hopping of three-hardcore bosons. In particular, the matrix elements of the second sub-/superdiagonal correspond in $H_{\mathrm{3hb}}$ to one-particle hopping terms
\begin{equation}
\label{eq:normalhop}
\ket{0}\ket{1} \leftrightarrow \ket{1}\ket{0}, \quad
\ket{1}\ket{1} \leftrightarrow \ket{2}\ket{0}, \quad
\ket{1}\ket{2} \leftrightarrow \ket{2}\ket{1};
\end{equation}
whereas those of the fourth sub-/superdiagonal correspond to a two-particle hopping
\begin{equation}
\label{eq:correlated}
\ket{0}\ket{2} \leftrightarrow \ket{2}\ket{0} .
\end{equation}
The relative importance of these two contributions can be modified just by tuning $|t_{\circ}|$, a factor which multiplies the second sub-/superdiagonal and is not present in the fourth one.

In the next sections we will try to use of the Hamiltonian in Eq.~(\ref{eq:bigmatrix}) to study 
interesting phenomena related to the presence of a three-hardcore constraint: 
we will focus on the realization of a quasi-condensate of pairs of particles and discuss the problems in realizing the two-dimensional Pfaffian state.
Before doing that, however, we will describe a setup providing us  an high external control on the hopping parameters
of the system, an inescapable requirement if we want to be able to realize the desired Hamiltonians.

\section{The Experimental Setup}\label{sec:experimental}

\begin{figure}[t]
\begin{center}
\includegraphics[width=0.9\columnwidth]{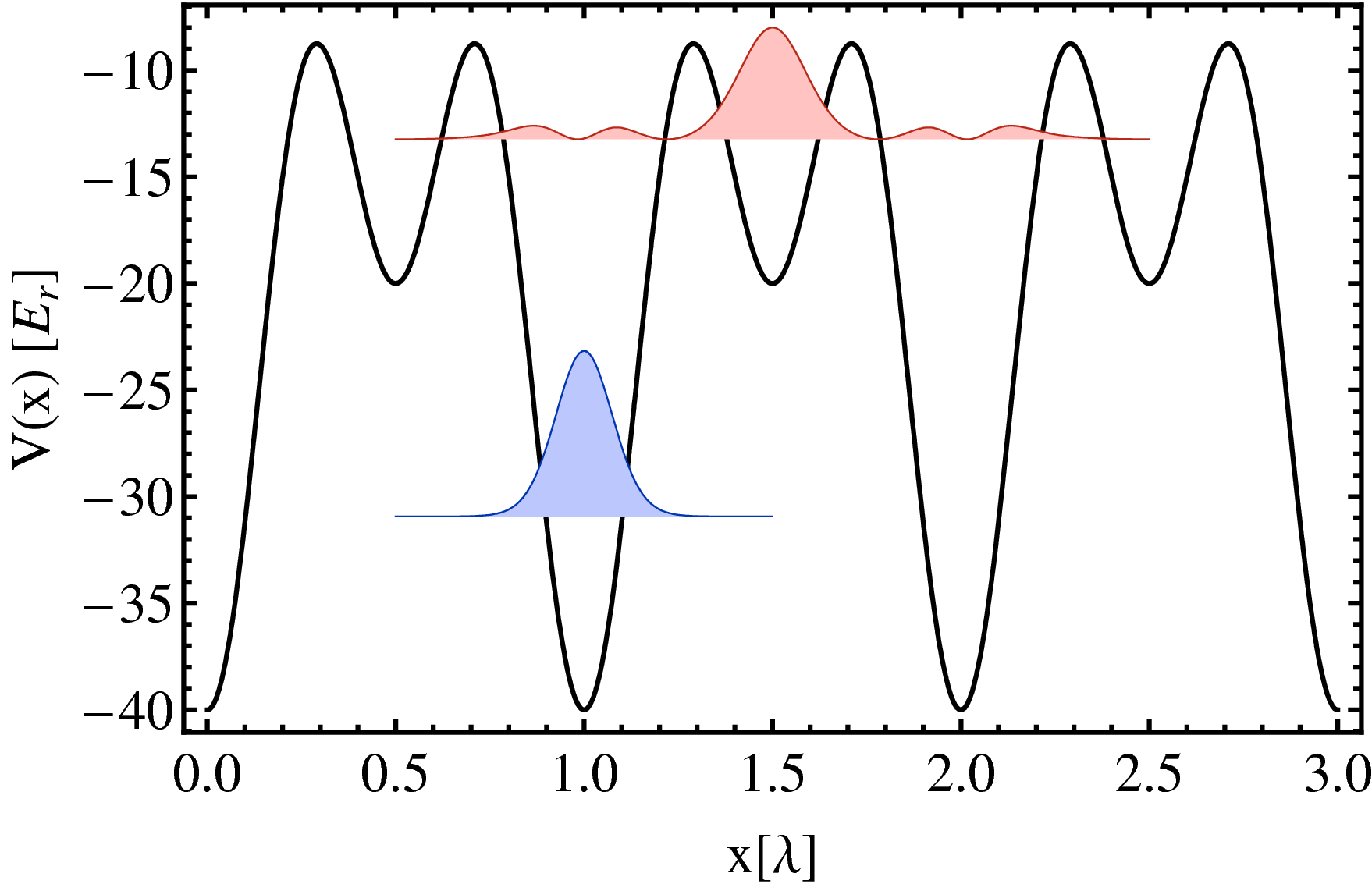}
\caption{Plot of the superlattice potential in Eq.~(\ref{eq:superlattice}) with $V_0 = 20 E_r$ and $\lambda = 1.0$, where $E_r = \hbar^2 k^2 /(2m)$. We show the exact wannier wavefunctions of the first and third band of the lattice, corresponding to the two bound states we will use in our proposal. }
\label{fig:superlattice}
\end{center}
\end{figure}

\begin{table}[b]
\begin{ruledtabular}
\begin{tabular}{|ccc|}
$E_{main}$ & $-30.9 E_r$ & $92.7$ kHz \\
$E_{sec}$ & $-13.2 E_r$ & $39.6$ kHz \\
$\delta = E_{sec} - E_{main}$ & $17.7 E_r$ & $53.1$ kHz\\
\hline 
$J_{main-main}$ & $-2.4 \cdot 10^{-4} E_r$ & $0.0$ Hz \\
$U_{0 main-main}$ & $1.8 E_r$ & $5.4$ kHz \\
\end{tabular}
\end{ruledtabular}
\caption{Numerical values of the parameters of the superlattice  in Eq.~(\ref{eq:superlattice}) with $V_0 = 20 E_r$ and $\lambda = 1.0$. We first list the energies of the lowest localized wannier functions of main and secondary minima and their energy difference; we then calculate the parameters of the main lattice, i.e. the hopping rate $J_{main-main}$ and the interaction energy $U_{0main-main}$.
}
\label{tab:parameters}
\end{table}

We now describe an optical lattice setup, consisting of a bichromatic spin-independent potential dressed with suitable optical transitions,
that once loaded with spin-1 atoms allow us to experimentally tune the spin-1 Hamiltonian in Eq.~(\ref{eq:ham:spin1:clean}).
Standard spin-independent optical lattices are indeed not suited for our purposes since they preserve SU(2) symmetry,
i.e. they naturally give rise to perfectly equal hopping rates $t_{\alpha}$ for all the spin orientations.
On the other hand, spin-dependent optical lattices~\cite{PhysRevA.70.033603, PhysRevLett.91.010407} suffer  heating problems and short lifetimes and are thus as well not optimal.
At a contrast, our setup allows for tuning independently the three hopping rates $t_{\alpha}$
without affecting the ``traditional'' quite long lifetimes.
In this Section we present a rather qualitative description, leaving quantitative analyses in  Appendix~\ref{app:superlattice}.
We start addressing the one-dimensional setup; the generalization to more dimensions is sketched at the end of this Section. 

\begin{figure}[t]
\begin{center}
\includegraphics[width=0.75\columnwidth]{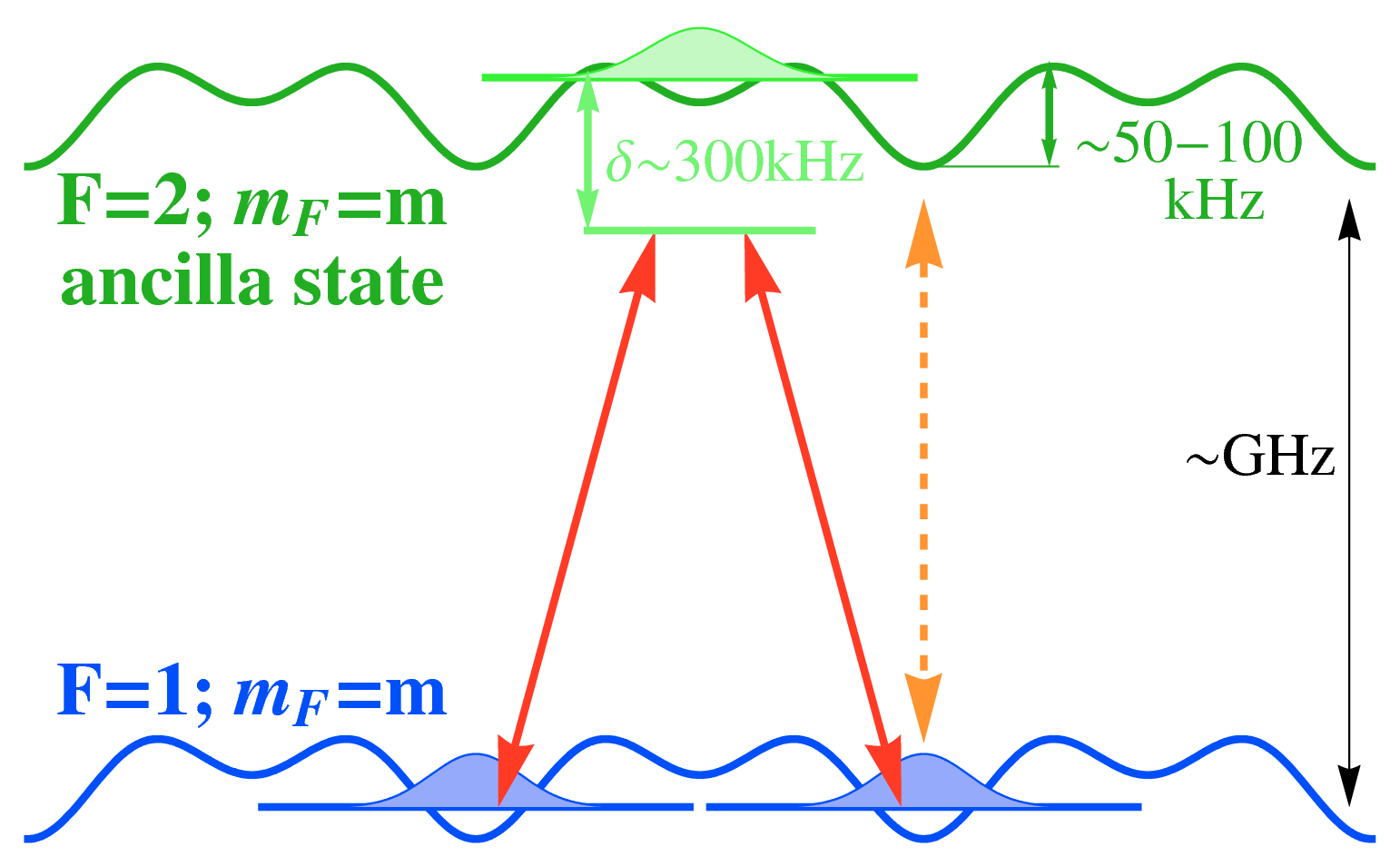}
\caption{
Sketch of the scheme we propose to induce hopping between the levels of the $F=1$ manifold, i.e. the adiabatic elimination of one $F=2$ state trapped in the intermediate minimum (red non-dashed arrows). Because of orthogonality properties of wannier functions, the coupling cannot be realised with microwave fields. Optical Raman transitions through an excited state carry non-negligible momentum and can therefore be a solution. In Appendix~\ref{app:superlattice} we  also discuss the effect of spurious couplings as those depicted with orange dashed arrows.}
\label{fig:transition}
\end{center}
\end{figure}

The alkaline atoms we propose to optically trap, $^{87}$Rb or $^{23}$Na, are characterized by a nuclear spin $I=3/2$ and thus 
by a ground state splitted into two hyperfine manifolds $F=1$ and $F=2$. 
If the laser originating the  lattice is detuned enough from the excited levels, the optical potential is insensitive to atomic spin properties and 
the eight ground hyperfine levels experience the same dipole potential~\cite{grimm}.
By combining two lasers with exactly commensurate frequencies 1:2 (obtainable by standard frequency-doublers) it is possible to create 
a bichromatic superlattice 
\begin{equation}
V(x) = - V_0 \left[ \cos^2 (k x) + \lambda \cos^2 (2 k x) \right]; \; \; \;  V_0, \lambda > 0 \; ,
\label{eq:superlattice}
\end{equation}
where $V_0$ is proportional to the overall light intensity and $\lambda$ describes the relative strength of the two lasers.
Tuning these two parameters, it is possible to create a configuration with main (deep) and secondary (shallow) minima,
as depicted in Fig.~\ref{fig:superlattice}. In particular, it is experimentally possible to cool a MI with atoms only in the main minima
just superimposing the second lattice after the cooling has already been done.
We propose to store all the physical information into the main minima of the $F=1$ manifold
whereas the other one will only provide auxiliary levels. Indeed, the idea is  to use the levels trapped in the secondary minima of the $F=2$ 
manifold   to provide intermediate ``bus'' states to be adiabatically eliminated, as pictorially explained in Fig.~\ref{fig:transition}.

The bound states energies and the corresponding localized Wannier wavefunctions were calculated solving the
single-particle Schr\"odinger equation with periodic boundary condition and manipulating the periodic solutions~\cite{KohnWannier}.
The overlap between them is exactly zero because they belong to different bands of the lattice spectrum,
making it impossible to implement couplings  via microwave fields that carry negligible momentum.
Our suggestion is to engineer them via optical Raman transitions, adiabatically eliminating a far excited state
like the one employed to create the trapping optical dipole potential and consequently transfering a momentum comparable
to the lattice inverse spacing. Provided a non-null intersection of the effective support for the Wannier wavefunctions, 
this scheme permits to have couplings like those of Fig.~\ref{fig:transition}.
We note that, because of the coherence properties of laser light, it is possible to give a complex phase to the effective hopping
and therefore to combine this setup with current proposals of artificial abelian and non-abelian gauge fields~\cite{1367-2630-5-1-356, Gerbier:Dalibard, MaciejSpielman}.
The results presented in this rticle were obtained by fixing the parameters  $V_0 = 20 E_r$ and $\lambda = 1.0$,
where $E_r = \hbar^2 k^2 /(2m)$ is the recoil energy of the long-wave-lattice;
some important parameters of this lattice are listed in Table~\ref{tab:parameters}. \\

\begin{figure}[t]
\begin{center}
\includegraphics[width=0.75\columnwidth]{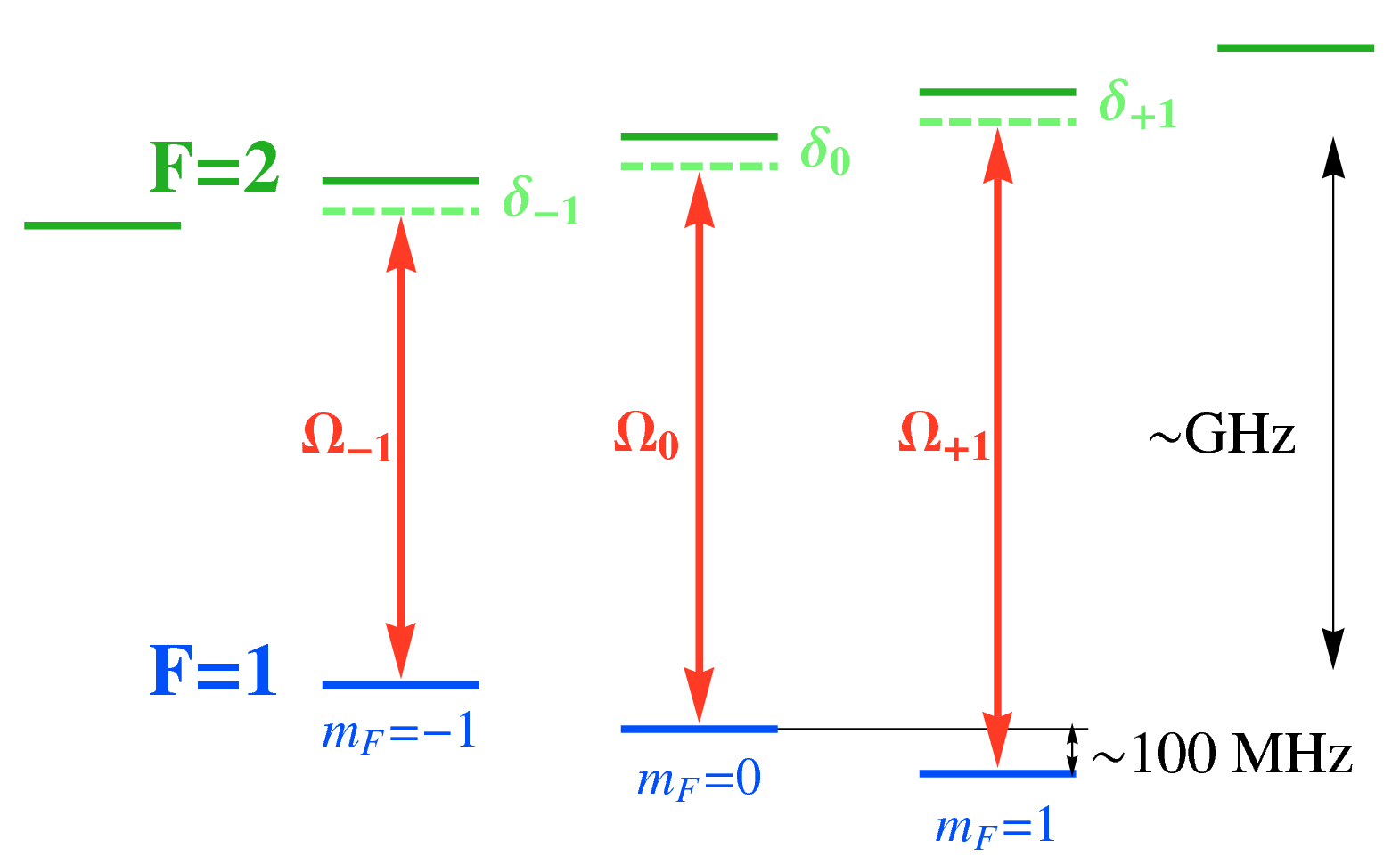}
\end{center}
\caption{Splittings of the levels of the $F=1$ and $F=2$ hyperfine manifolds in $^{87}$Rb due to an external magnetic field. 
The splitting between the two manifolds is not in scale. Red arrows describe the effective couplings we want to engineer via Raman transitions, $\delta$s and $\Omega$s are the effective parameters describing these transitions.}
\label{fig:splittings}
\end{figure}

We suggest to use the levels of the $\ket{F=2, m_F = m}$ as ancillas for the $\ket{F=1, m_F = m}$ states. 
The possibility of individually addressing each of the three transitions $\ket{1, m} \leftrightarrow \ket{2, m}$
is granted by an external magnetic field, which splits the levels according to the formulas:
\begin{eqnarray}
E_{1, m} = &  & -  g_F \, \mu_B \, B \; m +\Delta^{(2)}_{1,m} (\mu_B B)  \nonumber \\
E_{2, m} = & \Delta_{HF}  & +  g_F \, \mu_B \, B \; m + \Delta^{(2)}_{2,m} (\mu_B B)
\end{eqnarray}
where $\mu_B$ is the Bohr magneton, $g_F$ is the hyperfine Land\'e Factor and $\Delta_{HF}$ the hyperfine splitting (see Fig.~\ref{fig:splittings}). $\Delta^{(2)}_{F,m} (\mu_B B)$ denotes the second order Zeeman shift proportional to $(\mu_B B)^2$ of the magnetic sublevel $m_F = m$ of the spin-$F$ manifold, which in $^{87}$Rb has opposite signs for the two $F=1$ and $F=2$ manifolds.
The values for $^{87}$Rb are $\mu_B g_F = 0.7$MHz/G and $\Delta_{HF} = 6.8$GHz.
The transitions can be therefore detuned to a regime in which spurious effects can be safely neglected.
For example, if we consider $^{87}$Rb, even  weak magnetic fields of $10 - 100$ G can detune the three transitions of $15 - 150$ MHz.
A more quantitative and detailed discussion of these ideas is given in Appendix~\ref{app:superlattice},
where analytical and numerical arguments are provided in support of our proposal.
In particular, we discuss the effects of spurious couplings between other localised states of the two manifolds, represented in Fig.~\ref{fig:transition} by orange dashed lines, which prove not to affect the efficient population transfer between neighbouring sites.

As far as the generalization to more dimensions is concerned, we simply propose to apply the laser configuration originating the potential in Eq.~(\ref{eq:superlattice}) also in the other directions, labelled by $i$:
\begin{equation}
V(\mathbf x) \, = \,  - V_0 \sum_i  \left[ \cos^2 (k x_i) + \lambda \cos^2 (2 k x_i) \right]; \quad  V_0, \lambda > 0;
\end{equation}
An increased number of dimensions makes the structure of the relative minima more complicated,
however it is still possible to recognize a square or cubic geometry of main minima and an ``auxiliary'' lattice of secondary minima trapped
in the middle of links between the main ones. 
Other higher energy minima appear at the centers of faces and cubes, but they can be neglected due to their even higher energy offset.
Similarly to what done in one dimension, the system must be then dressed by optical lasers driving the nearest neighbor hopping.

Finally we notice that the other parameters appearing in Eq.~(\ref{eq:ham:spin1:clean}) can be experimentally tuned with
technologies standardly used in current optical lattice setups.
The interaction strengths $U_0$ and $U_2$ can be indeed modified via Feshbach resonances
of the scattering lengths $a_0$ and $a_2$~\cite{Bloch08,DalibardMWave}
whereas the energy offsets $\Delta_{\alpha}$ can be varied using the Zeeman effect or dressing the levels
with far-detuned microwave fields~\cite{PhysRevA.73.041602}. 

Before concluding this Section, we would like to remark that  we are suggesting to observe global magnetic properties
arising from the super-exchange effects of a spinorial Mott Insulator. Currently a lot of experimental efforts are devoted to this task,
mainly within the context of the Fermi-Hubbard model, the challenge being represented by severe temperature and entropy requirements~\cite{PhysRevLett.103.140401, PhysRevLett.104.180401, ketterle:temperature}.
Nonetheless, the huge quest going on makes us believe that such phenomena will be experimentally achievable in the next-future.
In the same spirit we intend also the superlattice setup, which at a first glance could seem rather intricated.

\section{Pairs Quasi-Condensation}\label{sec:pqc}

In this Section we study the first interesting model characterized by three-body infinite repulsion: we show that a dominating correlated hopping can drive a transition to a quasi-condensate of pairs (PQC), without the need of any two-body attraction. Moreover, we show that
substituting the three-body interaction with a two-body one the system becomes unstable towards collapse: this strictly links the PQC to the stabilizing effect of the three-body repulsion.

The experimental realisation of such a phase with the help of our setup is possible.
Besides the 3-body hardcore constraint, we already highlighted in Sec.~\ref{sec:spin1atoms} the simultaneous presence
of the usual single-particle hopping (\ref{eq:normalhop}) and of a correlated two-particle term (\ref{eq:correlated})
in the emerging Hamiltonian $H_{\mathrm{3hb}}$ of Eq.~(\ref{eq:bigmatrix}).
As already pointed out at the end of Sec.~\ref{sec:spin1atoms}, the relative strength of these two terms can be varied by just tuning the strength of $|t_{\circ  }|$: with the help of the setup discussed in Sec.~\ref{sec:experimental} it becomes feasible.
We explicitly discuss the realistic case of Eq.~(\ref{eq:bigmatrix}) where many spurious terms emerge but do not prevent PQC to be observed.

At zero temperature ($T=0$), very general theorems state that no long-range order can arise in 1D and Bose-Einstein condensation is
consequently ruled out; anyway, the presence of algebraical decays in the density matrix allows one to introduce the concept of
quasi-long-range order and quasi-condensation~\cite{Stringari}.
Beside the usual ``atomic'' quasi-condensate (AQC), characterized by quasi-long-range order of the one-particle density matrix
$\langle a_i^{\dagger}  a_{i+\Delta} \rangle$, it is possible to speak of  ``pairs'' quasi-condensate 
when the two-body  density matrix
$\langle a_i^{\dagger 2}  a^2_{i+ \Delta} \rangle$ still exhibits quasi-long-range order despite the exponential suppression of one-particle
correlations.

Even if mean-field calculations support the conjecture that PQC induced by correlated hopping is not a low-dimensionality phenomenon, the simplest setup for both experimental and numerical purposes is offered by a one-dimensional lattice.
On one side, a reduced number of Raman pairs of beams is required;
on the other, numerical simulations can be routinely done with the help of Density Matrix Renormalization Group
(DMRG)~\cite{white92,schoellwoeck}.
In this paper we made use of an open-source code (\texttt{www.dmrg.it}) with open boundary conditions (OBC) but in principle also
parabolic external potentials - closer to current experimental setups - lie within the possibilities of the method~\cite{schoellwoeck}.

Before starting our discussion, we report that the same paired phase has been also at the focus of Ref.~\cite{daley:040402} where the transition was driven
via two-body attractive interactions and the three-body hardcore constraint was effectively induced by strong dissipation channel.

\subsection{PQC Induced by Correlated Hopping}

\begin{figure}[t]
\begin{center}
\includegraphics[width=\columnwidth]{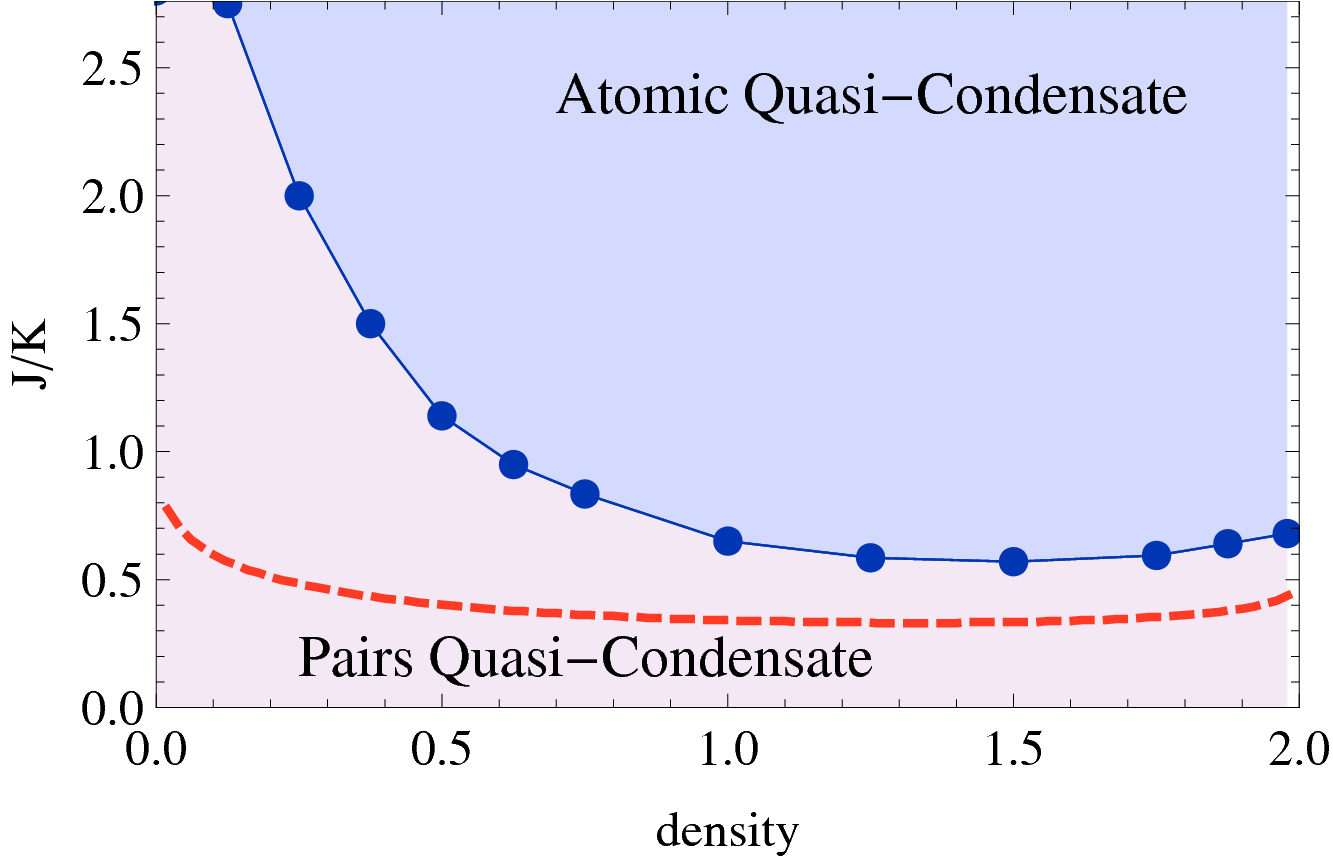}
\caption{Phase diagram of the one-dimensional model in Eq.~(\ref{eq:ehbc_and_corrhopp}) calculated with a DMRG algorithm. Two phases appear, characterized respectively by quasi-long-range order of the one-body density matrix (AQC) and by exponential decay of the one-body-density matrix and quasi-long-range order of the two body one (PQC). The red dashed line is the mean-field result obtained via Gutzwiller ansatz.}
\label{fig:PhaseDiag3bhc}
\end{center}
\end{figure}

In order to understand whether a system with dominating correlated hopping undergoes a phase transition to PQC, we start studying this simple ``blackboard'' Hamiltonian on a one-dimensional setup with $L$ sites:
\begin{equation}
H = -J \sum_{i} a^{\dagger}_i a_{i+1} - K \sum_i a^{\dagger 2}_i a^2_{i+1} + H.c.; \; \; \; (a_i^{\dagger})^3 = 0.
\label{eq:ehbc_and_corrhopp}
\end{equation}
The correspondent phase diagram, plotted in Fig.~\ref{fig:PhaseDiag3bhc}, displays a large region characterized by an exponential decay
of the particle-particle correlator $\langle a^{\dagger}_i a_j \rangle$ and by an algebraic decay of the the pair-pair one
$\langle a^{\dagger 2}_i a_j^2 \rangle$.
A mean-field analysis via Gutzwiller ansatz shows the presence of the same phase transition and thus supports the robustness of the effect
even in larger dimensions. It is also possible to Fourier transform both density matrices and analyze the finite-size scaling
of the population in the largest occupied  state, i.e. the lowest momentum one.
An algebraic growth $\sim L^{\alpha}$ is another signature of quasi-long-range order, and we find agreement between the two benchmarks.
In the following, we only look at the decay of the correlators.

\begin{figure}[t]
\begin{center}
\includegraphics[height=5.cm]{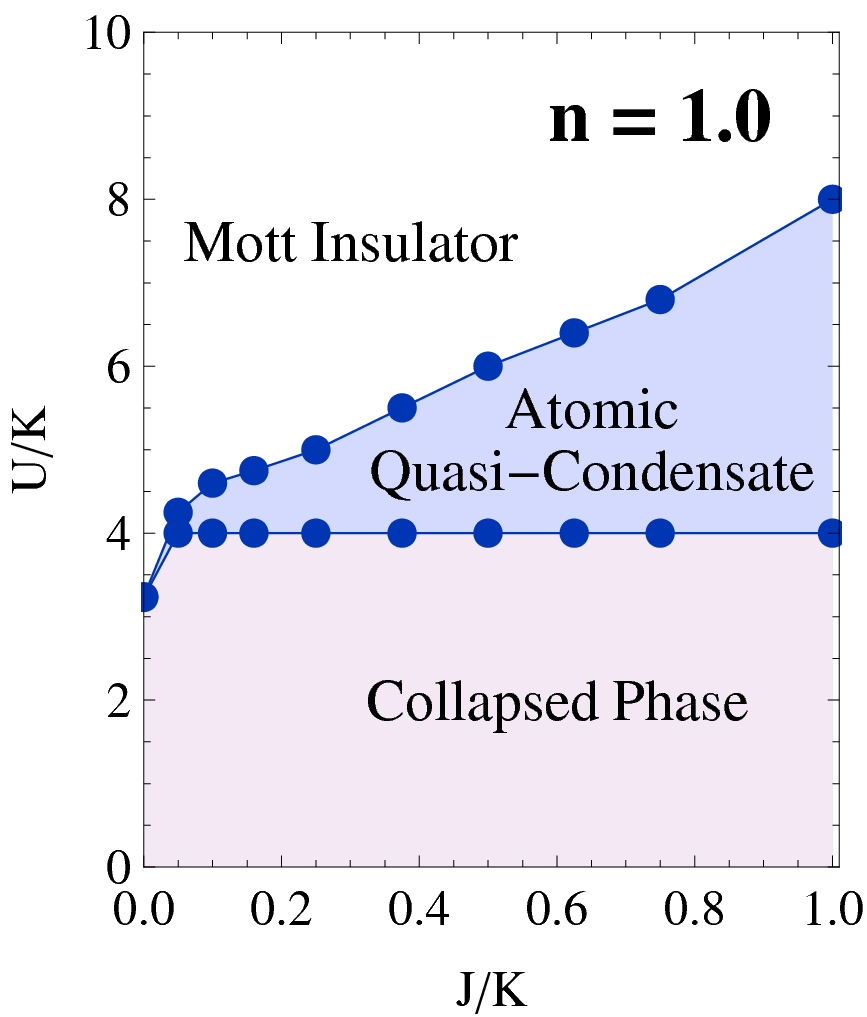}
\includegraphics[height=5.cm]{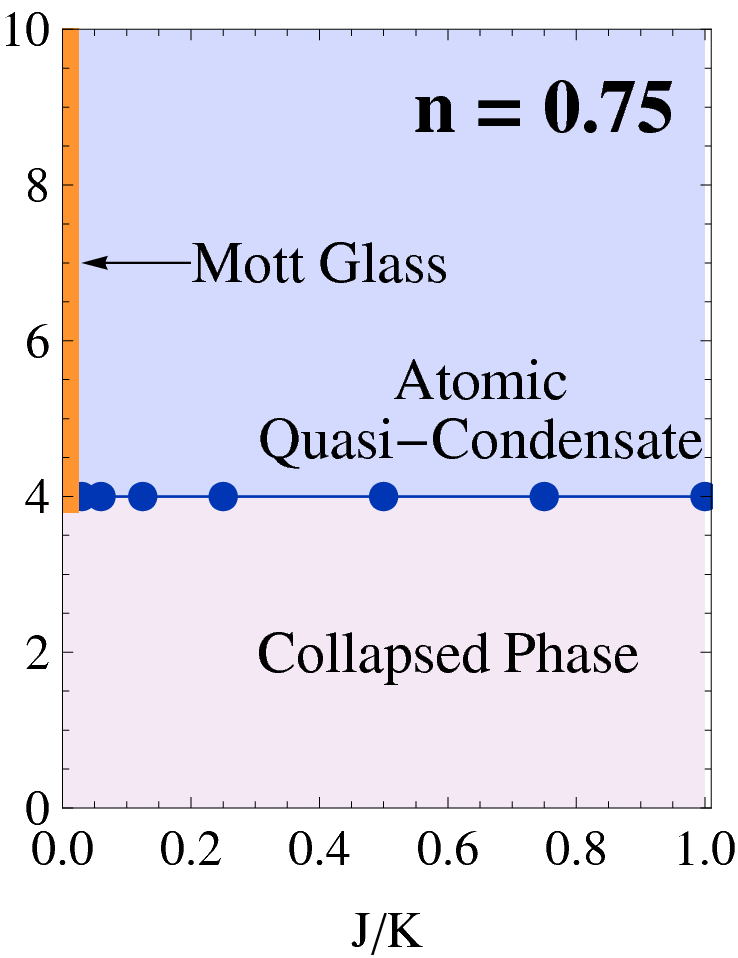}
\caption{Phase diagram of the system described by Eq.~(\ref{eq:fake:2body}) in which the three-body interactions have been substituted by two-body ones. Instead of a PQC a phase appears which is unstable towards collapse. At incommensurate filling and $J=0$ the ground state manifold is largely
degenerate and is spanned by Fock states with less than double local occupancies (Mott Glass).}
\label{fig:phdi:2body}
\end{center}
\end{figure}

We stress that the only presence of correlated hopping is not enough to create a PQC and that the stabilizing action of 
a three-body hardcore constraint is for this sake crucial. For example, we can  study a model in which we substitute the  constraint with two-body repulsions:
\begin{eqnarray}
H &=& -J \sum_{i} \left[a^{\dagger}_i a_{i+1} + H.c.\right] + \nonumber \\ 
&& - K \sum_i  \left[a^{\dagger 2}_i a^2_{i+1} + H.c. \right] + U \sum_i n_i (n_i - 1) \qquad 
\label{eq:fake:2body}
\end{eqnarray}
The relative phase diagram is plotted in Fig.~\ref{fig:phdi:2body} for two different densities, $n=1.0$ and $n=0.75$: 
when $K$ dominates a phase appears which in the thermodynamic limit is unstable towards collapse.
The instability is induced by the correlated hopping term and has been already observed in other numerical works~\cite{maria}; 
in Eq.~(\ref{eq:ehbc_and_corrhopp}) it was counterbalanced by the hardcore constraint.
The stabilizing action of strong two-body $U$ drives the system outside this instability region, but alas prevents
also double occupancies and then the desired PQC. In the limit of large $J/K$ the usual Bose-Hubbard physics
made of AQC and MI only is recovered (not shown in plots).

\subsection{PQC in Spin-1 Mott Insulators}

Coming back to the experimental setup proposed in Sec.~\ref{sec:spin1atoms}, we investigate whether it supports the pairs quasi-condensate.
In order to do that, we use a proper mapping~(\ref{eq:mapping}) $\mathsf W_{\mathrm{PQC}}$ with all the phases set to a same value $\varphi_{\alpha} = 0$.
The diagonal terms in Eq.~(\ref{eq:bigmatrix}) describe additional two-body and nearest-neighbors interactions that we cannot get rid of; 
therefore the
Hamiltonian in Eq.~(\ref{eq:ehbc_and_corrhopp}) cannot be easily recast and additional numerical simulations are needed to characterize our
approximation.

\begin{figure}[t]
\begin{center}
\includegraphics[height=5.cm]{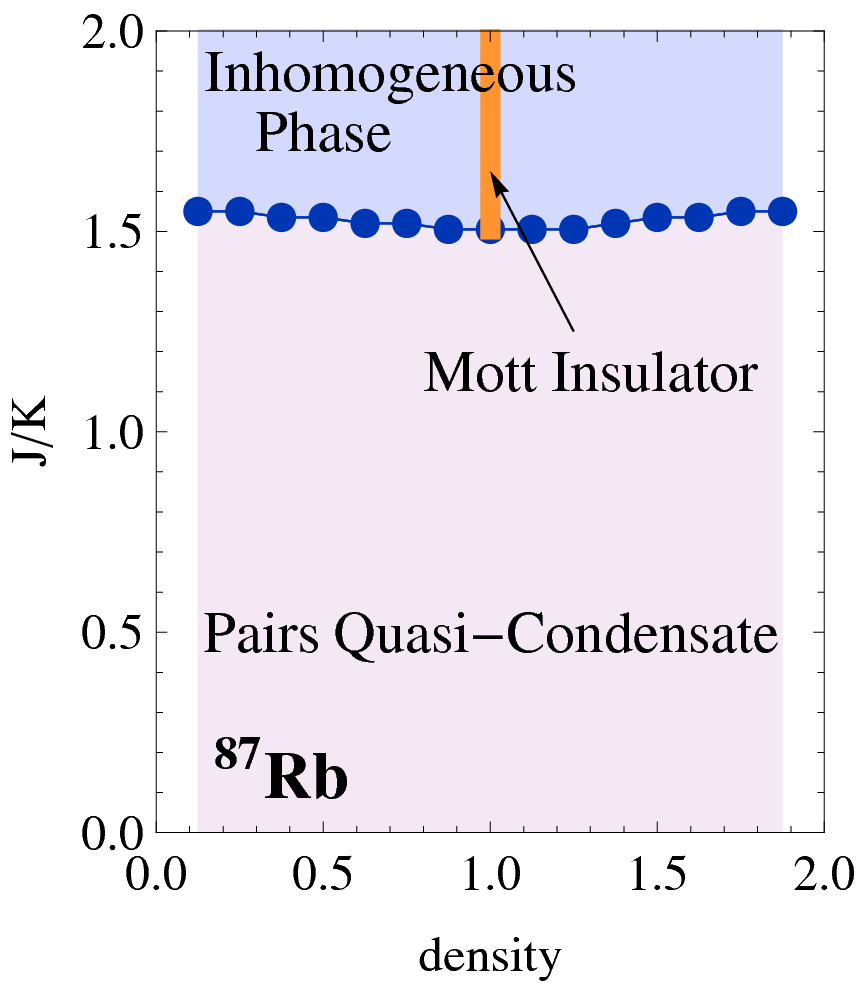}
\hspace{-0.5cm}
\includegraphics[width=0.5\columnwidth]{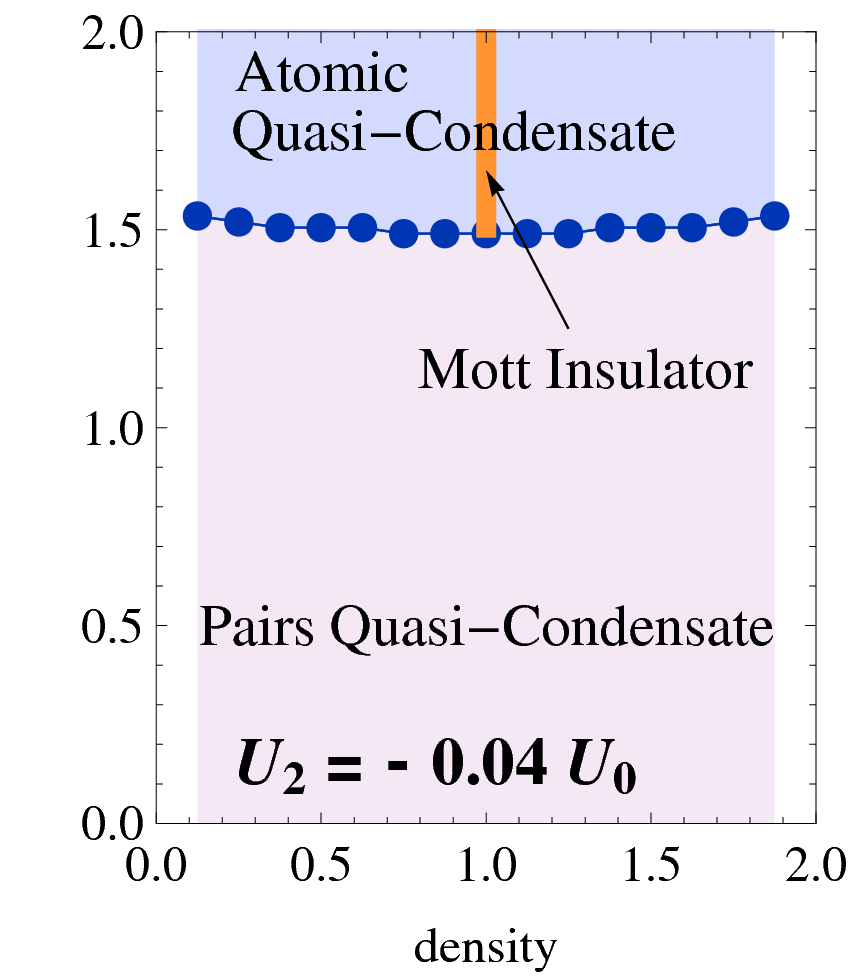}
\caption{
Phase diagrams of the
Hamiltonian $H_{\mathrm{3hbc}}$ realized with our proposal which approximates the model in Eq.~(\ref{eq:ehbc_and_corrhopp}).
The two plots are drawn for the realistic values of $U_2/U_0 = -0.005$ ($^{87}$Rb, left) 
and $U_2/U_0 = -0.04$ (right); the case $U_2/U_0 = 0.04$ ($^{23}$Na) is not shown since it is qualitatively equivalent to the case of $^{87}$Rb. On the left, even if a PQC phase appears, there are no signatures of AQC. Instead, at density $n=1.0$ we find a MI whereas at $n \neq 1.0$ an inhomogeneous phase appears (phase separation).
On the right, the phase diagram shows that even a small tuning of $U_2 / U_0$ from the atomic values let the AQC phase arise.}
\label{fig:phDi:realisticvalues}
\label{fig:PhaseDiag3bhc01}
\end{center}
\end{figure}

All the next plots share the same value of $t_{\pm }/U_0 = 0.1$ and of $\delta=0$, i.e. no relevant quadratic corrections to the linear Zeeman splitting.
With respect to the ratio $U_2/U_0$, we studied the cases of $^{87}$Rb ($-0.005$) and $^{23}$Na ($+0.04$) as well as the value $U_2/U_0 = -0.04$.
The phase diagrams, shown in Fig.~\ref{fig:phDi:realisticvalues}, have been studied varying the total density of three-hardcore bosons,
i.e. the magnetization of MI, and the ratio between single-particle and correlated hopping:
\begin{equation}
\frac JK = \frac{2 t_{\circ } t_{+1}}{U_0 + U_2} \frac{U_0 + U_2}{\sqrt 2 t_{+1} t_{-1}} = 10 \sqrt 2 \frac{t_{\circ}}{U_0}
\end{equation}

For the values of the two alkalis the system clearly exhibits a PQC phase, even if no clear signature of a AQC phase has not been found.
Instead an inhomogeneous phase appears characterized by phase separation between fillings $0-1$ or $1-2$.
However, a slight tuning of $U_2/U_0$ to $-0.04$ could already help the system recover all the interesting physics of the Hamiltonian in Eq.~(\ref{eq:ehbc_and_corrhopp}), as shown by the phase diagram in Fig.~\ref{fig:PhaseDiag3bhc01}, calculated for such value.
In this case the matrix elements of the second sub-/superdiagonal are still far from  being proportional to the exact values $\{1, \, \sqrt 2, \, \sqrt 2, \, 2 \}$. The phase transition is observable because the finely tuned values of the matrix elements are not important, as far as they roughly share the same order of magnitude and this last value is varied with respect to the correlated hopping parameter.

\begin{figure}[t]
\begin{center}
\includegraphics[width=0.85\columnwidth]{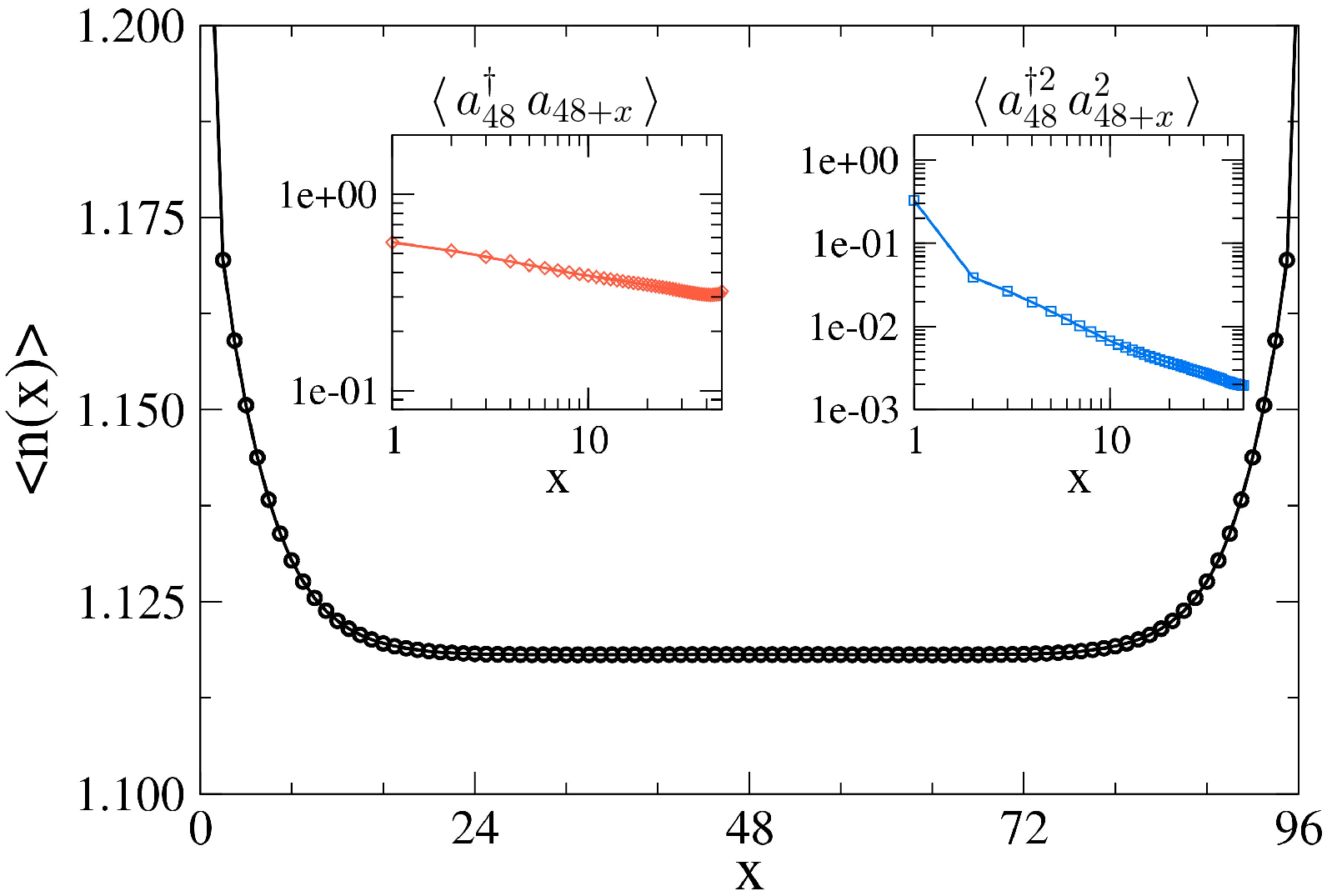}
\caption{Plot of the density profile $\langle n(x) \rangle$ of the ground state of the system for $U_2=-0.04 \,U_0$, $n=1.125$ and $J/K=1.54$. The two insets show the exponential decay of the particle-particle and pair-pair correlators  (log-log plots). These data allow us to identify the phase as a AQC.}
\label{fig:Plot_3bhc_U01_n125_JK170}
\end{center}
\end{figure}

\begin{figure}[t]
\begin{center}
\includegraphics[width=0.85\columnwidth]{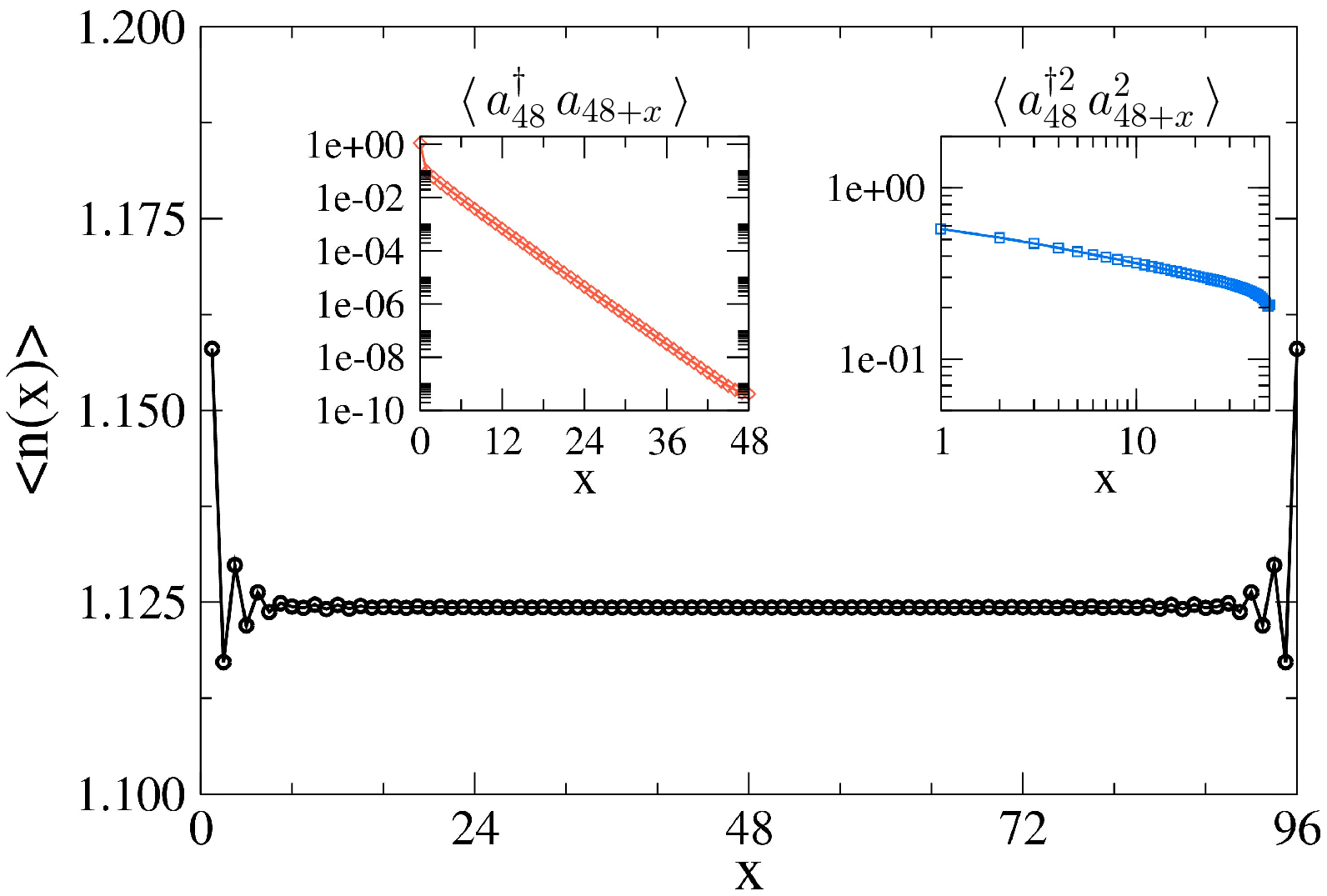}
\caption{Plot of the density profile $\langle n(x) \rangle$ of the ground state of the system for $U_2=-0.04 \,U_0$, $n=1.125$ and $J/K=1.43$. The two insets show the exponential decay of the particle-particle correlator (log plot) and the algebraic decay of the pair-pair correlator (log-log plot). The presence of this last quasi-long-range order allows us to identify the phase as a PQC.}
\label{fig:Plot_3bhc_U01_n125_JK127}
\end{center}
\end{figure}

The density profile and the spatial decays of the particle-particle and pair-pair correlators for the system at $n=1.125$ and $J/K=1.54$ (AQC) are shown in Fig.~\ref{fig:Plot_3bhc_U01_n125_JK170}, whereas in Fig.~\ref{fig:Plot_3bhc_U01_n125_JK127} the plots refers to $n=1.125$ and $J/K=1.43$ (PQC). The plots show quite clearly the presence of a region in the phase space where the pair-pair correlator exhibits quasi-long-range order whereas the particle-particle one is exponentially suppressed. The accurate definition of the phase border, requiring numerics on larger systems and finite size scalings, lies beyond the purposes of this article, which only aims in determining the presence of a PQC phase. However, simulations for large systems up to 240 sites show that the PQC phase is indeed stable and is not a finite-size effect.

For a more accurate theoretical treatment of the properties of the transition from QC to PQC we refer the interested reader to Ref.~\cite{arxiv:daley}. 
The applicability of such theoretical methods to our system must not be taken for granted because in our setup the phase transition is induced by correlated hopping whereas in the referred papers by two-body attractive interactions.

\subsection{Experimental Observation}

Since one of the key features of a quantum simulator is the possibility of observing  the quantum state that have been realised, we  now discuss how the two AQC and PQC phases could be detected with our setup. 
For this purpose, we translate into the spin language the correlators $\langle a^{\dagger}_ia_j\rangle$ and $\langle a^{\dagger 2}_i a_j^2 \rangle$ which we used to identify the two phases.
Once the mapping $\mathsf W_{\mathrm{PQC}}$ is considered, the three-body hardcore operators can be rewritten as follows:
\begin{equation}
a_i = \left( 1 + \frac{\sqrt2 - 1}{\sqrt2} S^z_i \right) S_i^+, \qquad
a_i^2 = \frac{1}{2}  S_i^{+ 2}.
\end{equation}
The pair-pair correlator assumes therefore a very simple expression, whereas the particle-particle one can  be written as:
\begin{eqnarray}
\langle a_i^{\dagger} a_j \rangle &=& 
\langle S_i^+  S_j^- \rangle + \nonumber \\
&+& \frac{\sqrt 2 - 1}{\sqrt 2}
\left[ \langle  S_i^z S_i^+  S_j^- \rangle +  \langle S_i^+  S_j^- S_j^z \rangle \right] + \nonumber \\
&+& \frac{3 - 2 \sqrt 2}{2} \langle S_i^z S_i^+  S_j^- S_j^z \rangle 
\end{eqnarray}

Numerical simulations show that the analysis of the decay of the spin correlators $ \langle S_i^+  S_j^- \rangle$ and $ \langle S_i^{+2}  S_j^{-2} \rangle $ leads to definition of the same phase boundary as before.
The possibility of using spin-spin correlators to identify the phases is experimentally relevant, because this is the most natural language to analyze 
the properties of a Spin-1 Mott Insulator.
Moreover, we think that the actual significative experimental efforts in order to develop techniques able to resolve the single sites of optical lattices~\cite{greiner:74,bloch_microscopy} will make the direct observation of the proposed correlators possible.

\section{Pfaffian Physics}\label{sec:pfaffian}

In this Section we move to the analysis of a second interesting many-body system characterized by three-body interactions: the Pfaffian wavefunction~\cite{Greiter1992567}. This state has been 
proposed in the context of the Quantum Hall Effect (QHE)~\cite{Ezawa00}
in order to describe the many-body electron liquid at fractional magnetic filling $\nu = 5/2$.
The interest in this wavefunction resides in the predicted property of supporting non-Abelian quasi-excitations~\cite{pfaffian, PhysRevB.54.16864}.

Here we deal with the bosonic version of the Pfaffian state ($\nu = 1$) and show that this wavefunction can be studied also in a lattice. Combining exact-diagonalization numerical approaches and some well-known benchmarks to test topological properties, we see that even at significative magnetic fields the ground state of the system features  non-trivial topological hallmarks. 
We then employ these tools to
discuss the possibility of using a spin-1 MI to realize such a wavefunction,
and underline some still present drawbacks in the recipe.

\subsection{Quantum Hall Effect on a Lattice}

We consider a two-dimensional setup with $N$ bosons with charge $q$ interacting via purely three-body repulsion (no two-body term)
in presence of an external uniform magnetic field with vector field $\mathbf A$.
The setup is pierced by a number of magnetic fluxes $N_{\Phi}$ equal to the number of particles $N$ (filling factor $\nu = 1$);
a typical length  $\ell = \sqrt{\hbar c  / q B }$ is induced in the system by the magnetic field itself.
The system is ruled by the following many-body Hamiltonian, 
in which we write the position of the particles with complex coordinates $z = (x+iy)/\ell$:
\begin{equation}
H_{\mathrm{Pf}} = \sum_i \frac{ [\mathbf p_i- \frac qc\mathbf A (z_i) ]^2 }{2m}
+
c_3 \hspace{-0.15cm} \sum_{i<j<k}   \hspace{-0.15cm} \delta( z_i - z_j) \delta( z_i - z_k).
\end{equation}
$c_3$, greater than zero, is the strength of the repulsion.
The single particle levels are arranged into a collection of degenerate manifolds, the Landau Levels (LL),
separated by a gap twice the cyclotron frequency $2 \hbar (q B / m c)$;
as long as the chemical potential is smaller than this separation, the particles will live only in the lowest LL
and will be characterized by wavefunctions analytical in $z$ (the exponent being the angular momentum).
Within this framework, the double-delta potential is properly regularized and
the ground state of the Hamiltonian is the Pfaffian wavefunction \cite{Greiter1992567}:
\begin{equation}
\Psi(z_1, ... z_N) \propto \text{Pf} \left( \frac{1}{z_i - z_j} \right) \  \prod_{i < j} (z_i - z_j) \  e^{- \sum_j |z_j|^2/2 }.
\end{equation}
The Slater determinant $\prod_{i < j} (z_i - z_j)$ would prevent the coincidence of two or more particles in the same spatial position;
the prefactor $\text{Pf} (1 / (z_i - z_j) )$, the Pfaffian (square root of the determinant) of the antisymmetric matrix with elements $A_{ij} = 1/(z_i - z_j)$,
enables the superposition of two bosons but still forbids that of three. 
With this construction, the wavefunction is forced to be the lowest angular momentum state in the intersection between Lowest Landau Level and
kernel of the three-body interaction.

In order to discuss the possibility of simulating the Pfaffian state with our proposal, we have first to discretize the system.
We take into account the presence of a three-body interaction with $c_3 \rightarrow \infty$
introducing the three-hardcore bosons operators $a$ and $a^{\dagger}$ satisfing $a^3 = 0$ and $a^{\dagger 3} = 0$.
The discrete version of a kinetik Hamiltonian with minimal coupling is:
\begin{equation}
\label{eq:ham_1}
 H_{\mathrm{Pf-lat}} = - J \sum_{<i,j>} e^{i \phi_{i,j}} \, a^{\dagger}_{i} a_j + H.c.; \quad 
( a_i^{\dagger} )^3  = 0.
\end{equation}
As in every discrete U(1) gauge theory, the magnetic field coupling to the positional degrees of freedom of the particles
is represented by a phase $\phi_{i,j} = 2 \pi /\Phi_0 \int_i^j \bm A \cdot d \bm l$,  where $\Phi_0 = h c /q$ is the quantum of flux. 

We already discussed in Sec.~\ref{sec:mapping} and~\ref{sec:spin1atoms} how to experimentally deal with the lattice version of three-hardcore bosons;
we also showed that our superlattice setup is compatible with the general theoretical idea of inducing a phase in the hopping
with an electro-magnetic running wave as in the pioneering proposal by Jaksch \emph{et al.}~\cite{1367-2630-5-1-356}. 
A flurry of theoretical proposals and experimental attempts to realize an artificial gauge field for neutral atoms have been going on for years
now~\cite{PhysRevA.70.041603, PhysRevLett.94.086803, juzeliunas:025602, gunter:011604,Gerbier:Dalibard, lin:130401, MaciejSpielman}
and have been further spurred by the breakthrough work by Lin \emph{et al.}~\cite{spielman:nature} that illustrated the experimental realization in a BEC.
Therefore we think that also this technological aspect of our proposal lies within the next-future possibilities and the only problem we are left with is 
whether it is possible to realize 
the model Hamiltonian~(\ref{eq:ham_1})
within the framework of Spin-1 Mott Insulators with Raman superlattice dressings.

\subsection{Topological Properties as a Benchmark}

Before discussing the simulation of Hamiltonian~(\ref{eq:ham_1}), we investigate to which extent transposing the physical system onto a discrete lattice modifies the nature of the many-body state. The problem arises from the competition of two typical lengths, the magnetic one $\ell$ and the lattice constant $a$.
In the small magnetic field limit $l \gg a$ (or dilute limit, since the constraint $N = N_{\Phi}$ must hold),  we expect the system to be insensitive to the discrete nature of the space. On the other side, an analysis of what happens when the magnetic field (and the particle density as well) increases is needed to test the robustness of a fully discrete version of the Pfaffian wavefunction.

The characterization of QHE wavefunctions transposed from continuum systems (usually two-dimensional strongly interacting electrons) to discrete optical lattices is a problem that has already been faced in the literature \cite{hafezi:023613, refId, moller:105303}. 
Here we follow the standard approach. We perform an exact diagonalization of the system with periodic boundary conditions (PBC).
Three marks are used to test the genuine Pfaffian nature of the numerical ground state (see Table \ref{table:numerical} for their values in this case):
\begin{enumerate}
\item the agreement between the degeneracy of the discrete numerical and continuum analytical ground manifolds (the Pfaffian wavefunction has been generalised on a torus first in Ref.\cite{Greiter1992567});
\item a significative overlap of the discrete numerical wavefunctions with the continuum analytical ones;
\item the agreement between the Chern number (CN)~\cite{    JPSJ.73.2604, JPSJ.74.1374} of the discrete numerical and continuum analytical ground manifolds.
\end{enumerate}

We stress that the three-fold degeneracy of the pfaffian ground state is not of topological nature, and is strictly connected to the properties of the Jacobi theta functions, which are used to generalized on the torus some QHE states~\cite{Greiter1992567}.

Chern Numbers probe the topological properties of the system testing its sensibility towards the twist of the boundary conditions, expressed by two parameters $(\theta_x, \theta_y) \in [0, 2 \pi) \times [0, 2 \pi)$. We give here the expression of the first CN for the simple case of non-degenerate ground state, whereas for more dimensions we refer to Refs.~\cite{JPSJ.73.2604, hafezi:023613}:
\begin{equation}
C = \frac{1}{2 \pi} \int d \theta_x d\theta_y \left[
\partial_{\theta_x} A_y(\theta_1, \theta_2) -
\partial_{\theta_y} A_x(\theta_1, \theta_2)
\right]
\end{equation}
where $A_i = \langle \Psi(\theta_x, \theta_y) | \frac{\partial }{\partial \theta_i} | \Psi (\theta_x, \theta_y) \rangle$ and $\ket{ \Psi (\theta_x, \theta_y)}$ is the ground state with boundary conditions $(\theta_x, \theta_y)$.
This integer quantity is indeed related to the theory of topological invariants in the context of the Berry connection. CN are increasingly used in condensed-matter theory since the discovery that the quantized properties of the anomalous QHE resistivity could be studied within such framework~\cite{PhysRevLett.49.405}. We calculate the CN with the method provided by Hatsugai~\cite{JPSJ.73.2604} which avoids any explicit numerical differentiation 
and connects the CN evaluation to the number of vortices displayed by a proper auxiliary field $\Omega(\theta_x, \theta_y)$.
The consequent integer character of the CN constitutes a further reason of its extensive use, since it provides a reliable yes-no benchmark more immediate than the wavefunction overlap ranging in $[0, 1]$.

\begin{figure}[t]
\begin{center}
\includegraphics[width=0.85\columnwidth]{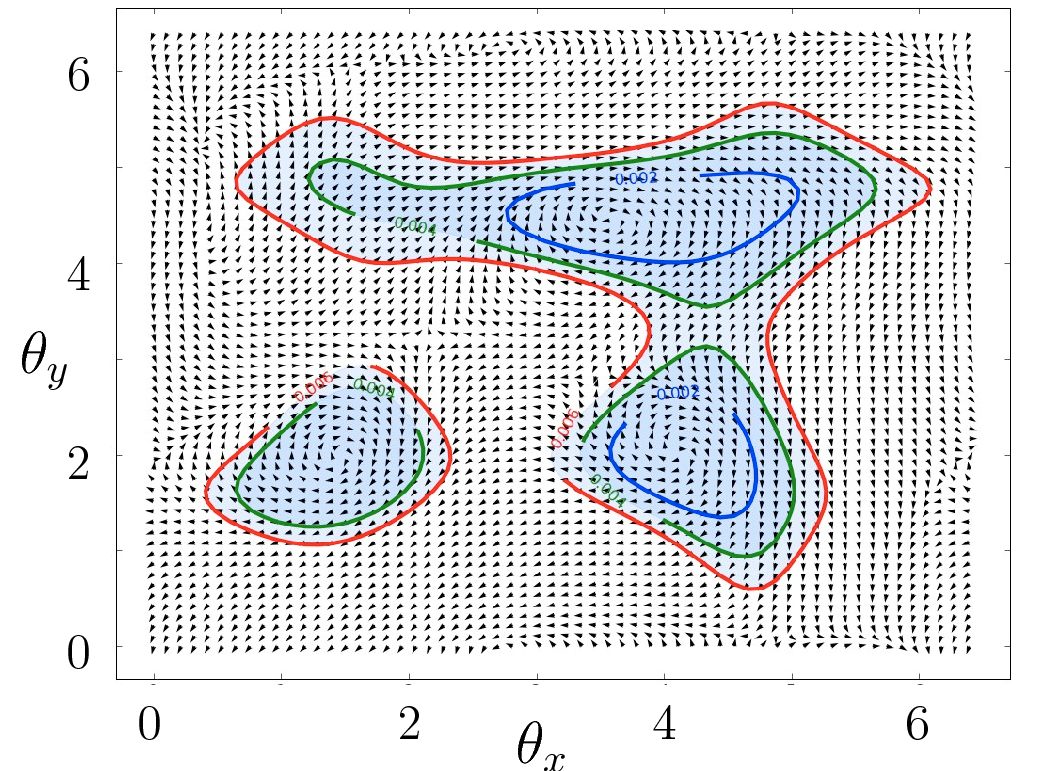}
\caption{Plot of the auxiliary field $\Omega(\theta_x, \theta_y)$ for the system with Hamiltonian (\ref{eq:ham_1}). The parameters of the system are those in Table~\ref{table:numerical}. The three highlighted vortices mean that the CN of the system is equal to 3. Vortices must be searched in regions where the main chosen gauge has minima: the contour lines and the shading highlight such part of the parameter space.
The definition of the field $\Omega$ and the way it can be computed are discussed extensively in Ref.~\cite{JPSJ.73.2604,hafezi:023613,refId}, to which we refer the interested reader.}
\end{center}
\end{figure}

Within the uncertainty given by working with small systems without accessing the thermodynamic limit,
we can at least affirm that our results are compatible with the presence of an incompressible liquid with a degenerate ground state
on the torus at $\ell \sim 0.8 \, a$.
Moreover, they also present significative signatures that the nature of the system should be strictly linked to that of the Pfaffian state.

\subsection{Tentatives Towards the Clean Model}

The previous Section shows that if we were able to implement Hamiltonian~(\ref{eq:ham_1}) we would access the intriguing physics of Pfaffian with our 
quantum simulator.
The Hamiltonian for effective bosons obtained in Eq.~(\ref{eq:bigmatrix}) has to be then compared with the link version of Eq.~(\ref{eq:ham_1})
\begin{displaymath}
H_{\mathrm{disc}} = \hspace{-0.1cm}-J
\hspace{-0.1cm}
\left(
\begin{array}{ccccccccc}
\squeezetable
0 & \hspace{-0.2cm} 0 & \hspace{-0.55cm} 0 & \hspace{-0.45cm} 0 & \hspace{-0.45cm} 0 & \hspace{-0.55cm} 0 & \hspace{-0.65cm} 0 & \hspace{-0.55cm} 0 & \hspace{-0.25cm} 0 \\
0 & \hspace{-0.2cm} 0 & \hspace{-0.55cm}0 & \hspace{-0.45cm} e^{i \phi_{i,j}} & \hspace{-0.45cm} 0 & \hspace{-0.55cm} 0 & \hspace{-0.65cm} 0 & \hspace{-0.55cm} 0 & \hspace{-0.25cm} 0 \\
0 & \hspace{-0.2cm} 0 & \hspace{-0.55cm} 0 & \hspace{-0.45cm} 0 & \hspace{-0.45cm} \sqrt 2 e^{i \phi_{i,j}} & \hspace{-0.55cm} 0 & \hspace{-0.65cm} \heartsuit & \hspace{-0.55cm} 0 & \hspace{-0.25cm} 0 \\
0 & \hspace{-0.15cm} e^{- i \phi_{i,j}} & \hspace{-0.55cm} 0 & \hspace{-0.45cm} 0 & \hspace{-0.45cm} 0 & \hspace{-0.55cm} 0 & \hspace{-0.65cm} 0 & \hspace{-0.55cm} 0 & \hspace{-0.25cm} 0 \\
0 & \hspace{-0.2cm} 0 & \hspace{-0.55cm} \sqrt 2 e^{- i \phi_{i,j}} & \hspace{-0.45cm} 0 & \hspace{-0.45cm} \spadesuit & \hspace{-0.55cm} 0 & \hspace{-0.65cm} \sqrt 2 e^{i \phi_{i,j}} & \hspace{-0.55cm} 0 & \hspace{-0.25cm} 0 \\
0 & \hspace{-0.2cm} 0 & \hspace{-0.55cm} 0 & \hspace{-0.45cm} 0 & \hspace{-0.45cm} 0 & \hspace{-0.55cm} 0 & \hspace{-0.65cm} 0 & \hspace{-0.65cm} 2 e^{i \phi_{i,j}} & \hspace{-0.25cm} 0 \\
0 & \hspace{-0.2cm} 0 & \hspace{-0.55cm} \heartsuit & \hspace{-0.45cm} 0 & \hspace{-0.45cm} \sqrt 2 e^{- i \phi_{i,j}} & \hspace{-0.55cm} 0 & \hspace{-0.65cm} 0 & \hspace{-0.55cm} 0 & \hspace{-0.25cm} 0 \\
0 & \hspace{-0.2cm} 0 & \hspace{-0.55cm} 0 & \hspace{-0.45cm} 0 & \hspace{-0.45cm} 0 & \hspace{-0.55cm} 2 e^{- i \phi_{i,j}} & \hspace{-0.65cm} 0 & \hspace{-0.55cm} 0 & \hspace{-0.25cm} 0 \\
0 & \hspace{-0.2cm} 0 & \hspace{-0.55cm} 0 & \hspace{-0.45cm} 0 & \hspace{-0.45cm} 0 & \hspace{-0.55cm} 0 & \hspace{-0.65cm} 0 & \hspace{-0.55cm} 0 & \hspace{-0.25cm} 0 
\end{array}
\right)
\end{displaymath}
where the graphic symbols highlight some terms of Eq.~(\ref{eq:bigmatrix}) which are not present in (\ref{eq:ham_1}).

\begin{table}[t]
\begin{ruledtabular}
\begin{tabular}{cccc|ccc|c}
$N$ & $N_{\Phi}$ & $L_x \times L_y$ & $\ell / a$ & degen. & overlap & CN & dim$\mathcal H$ \\
\hline
$4$ & $4$ & $4 \times 4$ & $\sim 0.8$ & $3$ & $78\%$ & $3$ & $3620$ 
\end{tabular}
\end{ruledtabular}
\caption{Exact diagonalization study on a torus of the many-body ground state of the system described by the Hamiltonian in Eq.\ref{eq:ham_1}. The degeneracy and the Chern number of the ground manifold in the continuum case are respectively $3$ and $3$. As discussed in Appendix~\ref{app:Flux}, the presence of magnetic fields strongly constraints the dimension of the torus to be simulated; the next size would be $5 \times 5$, with an Hilbert dimension $110$k.}
\label{table:numerical}
\end{table}

One of the problems is related to the presence in $\heartsuit$ of the correlated hopping term~(\ref{eq:correlated}),
which is not comprised by the the QHE model. Therefore, we tried to study the model in the regime: $|t_{\circ}| \gg |t_+|, |t_-|$, which decreases the relevance of correlated hopping. In this case we use a mapping $\mathsf W_{\mathrm{PF}}$ characterized by the phases: $\{ \varphi_- = 0; \varphi_{\circ} = 0; \varphi_+ = \pi  \}$ and set the various parameters to the values listed in Table~\ref{table:1parameters}. This sets the second sub-/superdiagonal to be approximately proportional to $\{ 1; \sqrt 2; \sqrt 2; 2\}$.
Unfortunately, this tunes only eight of the terms of the diagonal to an approximate same value: the central one $\spadesuit$ is significantly different from the others, leading to a completely different model with an effective nearest neighbours interaction. Moreover, this method has the general disadvantage that the effective hopping rate $J$ would be proportional to $|t_{\circ} t_-| / (U_0 + U_2)$ and therefore require temperatures even lower than the pure super-exchange effect $|t_{\circ}|^2 / (U_0 + U_2)$.

As an alternative, we abandon the attempt to exactly recover the model in Eq.~(\ref{eq:ham_1}) and try instead to 
realize a similar system whose ground state is characterized by the same benchmarks of the Pfaffian wavefunction,
i.e. the same degeneracy on the torus and the same Chern number. 
At low density, the number of global Fock states with more than two particles on one link is lower than that of the other states.
Thus, we expect that matrix elements of the link Hamiltonian connecting states with more than two particles per link 
do not play a relevant role in the global dynamics; 
even sensible deviations of such terms from the exact values should not change too much the properties of the ground state. 
Hence, we investigated sets of parameters which could put all the ``noise'' on such matrix elements.
We consider the same $4 \times 4$ system as before at density $\rho = 1/4$ and magnetic field $N_{\Phi} = 4$, which we can numerically analyze, but the next considerations could also be generalized to systems with smaller magnetic fields (or more dilute).
Unfortunately, even this turned out to be impossible. We tried to combine a tomographic analysis of the Pfaffian wavefunction with the tuning of all the matrix elements of the link Hamiltonian connecting states with less than three particles. However the numerical simulation of these Hamiltonians gave always as result non-degenerate ground states characterized by no topological properties, i.e. a Chern number equal to zero~\cite{JPSJ.74.1374}.\\

\begin{table}[t]
\begin{ruledtabular}
\begin{tabular}{c}
\begin{tabular}{cc}
$U_2 = \sqrt 2 / (2 \sqrt 2 +3) U_0 \sim 0.24 U_0 \quad$ & $\quad   \delta = - 2 |t_{\circ}|^2 / (U_0 + U_2)$ \\
\end{tabular}\\
\hline 
\begin{tabular}{ccc}
$t_{\circ} = 0.1 U_0 \quad$ & $ \quad t_- = 0.1 t_{\circ} e^{i \vartheta} \quad$ & $\quad t_+ = 2 t_-^*$  \\
\end{tabular}
\end{tabular}
\end{ruledtabular}
\caption{Set of parameters used together with the mapping $\mathsf W_{\mathrm{PF}}$ to recover the model in Eq.~\ref{eq:ham_1}. We stress that it is possible to give to the phase $\vartheta$ the space dependence which characterizes $\phi_{i,j}$.}
\label{table:1parameters}
\end{table}

Unfortunately we were then not able to find neither a way to get the Hamiltonian in Eq.~(\ref{eq:ham_1})
nor to realize a similar Hamiltonian whose ground state was three-fold degenerate and characterized by a Chern Number equal to three.
Ergo we think that, alas, the Pfaffian wavefunction cannot be readily implemented with the help of a quantum simulator based merely on the
ingredients described in this work. It might be nonetheless the case that, adding to the proposed setup  some furher trick or ancillary system,
it becomes feasible.

\section{Conclusions}\label{sec:conclusions}

In this article we have discussed two many-body examples of systems characterized by a three-body infinite contact repulsion. In the former we studied a one-dimensional phase characterized by quasi-long-range order induced by correlated hopping, whereas in the latter we have examinated the stability of a discrete bosonic Pfaffian wavefunction in a non-dilute limit.

Moreover, we suggest to experimentally realise such phases with the help of optical lattices and spin-1 atoms. Our proposal to simulate three-body infinte repulsion relies on a local mapping between the dynamics of a spin-1 MI and that of emerging bosons characterized by such interaction. Numerical calculations support the experimental feasibility of the former setup, whereas the latter it seems that further control parameters are still needed.

A crucial point of this paper is the extensive description of a bichromatic optical superlattice which could allow the realisation of rather general hopping operators for spin gases in optical lattices. 
We describe the case of spin-preserving hopping rates and show that it is possible with laser assisted tunneling to break the SU(2) symmetry. Moreover, in a next work we will show that an additional staggering of the lattice can in principle give access to a more general class of hopping operators operators containing even spin-flipping terms.
We hope with this to open a route towards theoretical studies of interesting models or towards the experimental realisation of exotic spin models.

\section{Acknowledgements}

We gratefully acknowledge fruitful discussions with U. Schneider, with whom the superlattice setup was actually conceived.
We also thank M. Aguado, N.R. Cooper, A.J. Daley, F. Gerbier, M. Hafezi, B. Horstmann, A. Muramatsu, M. Roncaglia,
T. Roscilde and D. Rossini for stimulating conversations concerning this work.
The research leading to these results has received funding from the European Community's Seventh Framework Programme (FP7/2007-2013) under grant agreement n¡ 247687 (Integrating Project AQUTE).
M.L. aknowledges the Spanish MICINN Grants (FIS2008-00784 and QOIT), EU STREP (NAMEQUAM), ERC Advanced Grants (QUAGATUA)
and the support of the Humboldt Foundation.
This work has been developed by using the DMRG code released within the ``Powder with Power'' project (www.dmrg.it).

\appendix

\section{External Control of the Hopping Rate via Superlattices \label{app:superlattice}}

In this Appendix we give a quantitative analysis of the discussion of Sec.~\ref{sec:experimental}, in which we suggested to use
superlattices in order to externally and independently control the hopping of the three spin species.

We start discussing the explicit expression of the coupling realised with an optical Raman transition between two different hyperfine levels of the ground state $L=0$ via elimination of the manifold of excited states $L=1$, where $L$ is the electronic angular momentum. Atomic levels are addressed with the  notation $\ket{L, \alpha, k}$, where $\alpha$ labels the hyperfine degrees of freedom and $k$ are the quantum numbers of the center-of-mass wavefunction. The Raman coupling between two states $\ket{0 \, \alpha \, k}$ and $\ket{0 \, \alpha' \, k'}$ is:
\begin{eqnarray}
\tilde{\Omega}_{\alpha' k'; \alpha k} (t) = - \frac 12\sum_{| 1\,\beta \, q\rangle} \bra{k'} e^{-i \mathbf{p}_2 \cdot \mathbf x} \ket{q} \bra{q} e^{i \mathbf p_1 \cdot \mathbf x} \ket{k} \cdot \nonumber \\
\cdot \; c_{2 \alpha' \beta}^* \, \|\mu \|_2^* \, E^*_2 \, E_1 \, \|\mu \|_1 \, c_{1 \alpha \beta} \; \cdot \; e^{- i (\omega_1 - \omega_2)t} \; \cdot \nonumber \\
\cdot \left(
 \frac{1}{E_{1 \beta q} - E_{0 \alpha k} - \hbar \omega_1}
+ \frac{1}{E_{1 \beta q} - E_{0 \alpha' k'} - \hbar \omega_2}
\right) 
\label{eq:longRaman}
\end{eqnarray}
where $E_{L \alpha k}$ is the energy of the level  $\ket{L, \alpha, k}$ and
$\omega_i$ and $\mathbf p_i$ are the energy and momentum of the $i$-th laser. The coupling realised by the $i$-th laser between the internal atomic states $\ket{0 \alpha}$ and $\ket{1 \beta}$ is described by $\|\mu \|_i$, $E_i$ and $c_{i \alpha \beta}$  according to the notation of Ref.~\cite{grimm}.
The sum over the excited states is limited to  the first excited manifold because we consider lasers far-detuned from higher excited levels. 
In the case of a spin-independent lattice, lasers can be detuned from the first excited manifold $L=1$ of even some tens of THz: in this case the expression in Eq.~(\ref{eq:longRaman}) can be simplified. Indeed, the energy differences at the denominators depend only slightly on the internal structure of the levels (they can differ at most  for some GHz): once $E_{1 \beta q} - E_{0 \alpha' k'}$ is substituted with the $0$-th order energy difference between excited and ground states $\Delta E_{10}$, we can write:\\
\begin{eqnarray}
\tilde{\Omega}_{\alpha' k'; \alpha k} (t)  = \phantom{ciao ciao ciao ciao ciao ciao ciao ciao ciao} \nonumber \\
 = - \frac 12 
\left(
 \frac{\bra{k'} e^{-i (\mathbf p_2 - \mathbf p_1) \cdot \mathbf x} \ket{k}}{\Delta E_{10} - \hbar \omega_1}
+ \frac{\bra{k'} e^{-i (\mathbf p_2 - \mathbf p_1) \cdot \mathbf x} \ket{k}}{\Delta E_{10} - \hbar \omega_2}
\right) 
\cdot \nonumber \\
 \cdot 
\sum_{\beta} c_{2 \alpha' \beta}^* \, \|\mu \|_2^* \, E^*_2 \, E_1 \, \|\mu \|_1 \, c_{1 \alpha \beta} \; 
\; e^{- i (\omega_1 - \omega_2)t} =  \nonumber \\ 
= S_{k'k} \, \Omega_{\alpha' \alpha} \, e^{- i \omega t} \phantom{ciao ciao ciao ciao ciao ciao ciao c}
\label{eq:shortRaman}
\end{eqnarray}
In this last expression $\omega = \omega_1 - \omega_2$, $S_{k'k} = \bra{k'} e^{-i (\mathbf p_2 - \mathbf p_1) \cdot \mathbf x} \ket{k}$ whereas $\Omega_{\alpha' \alpha}$ comprises all the remaining terms.
The very simplified expression for the center-of-mass part of the coupling $S_{k'k}$ comes from the substitution of $\sum_q \ket{q} \bra{q}$ with the identity on the center-of-mass Hilbert space.

Taking advantage of Eq.~\ref{eq:shortRaman} specified to the setup described in Sec.~\ref{sec:experimental}, we now discuss the possibility of transferring population between two $F=1$ neighbouring sites via adiabatic elimination of an $F=2$ state trapped in the middle.

\begin{figure}[t]
\begin{center}
\includegraphics[width=0.9\columnwidth]{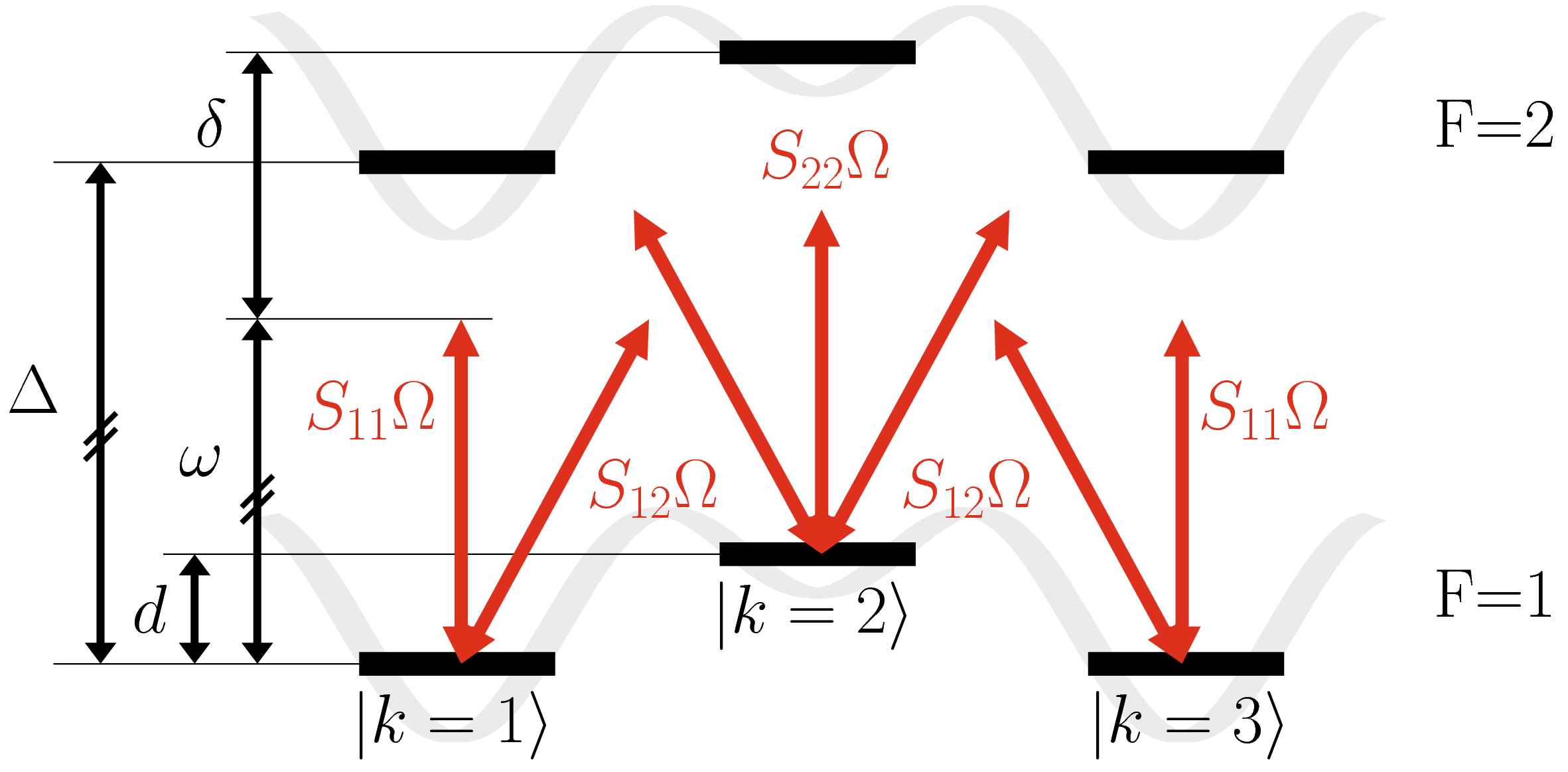}
\caption{The ``6-levels model'' used to study the coupling of different hyperfine levels with one Raman transition. Levels are labelled by two quantum numbers, $F$ and $k$. Energies are not in scale; the orders of magnitude of the parameters are the following: $d \sim 10\div 100$ kHz, $\delta \sim 100 \div 500$ kHz and $\Delta \sim 1\div 10$ GHz. We propose to adiabatically eliminate the upper manifold and to study the dynamics of the lowest one with an effective Hamiltonian $H_{pert}$.}
\label{fig:link}
\end{center}
\end{figure}

We consider two states with the same magnetic quantum number $m_F$, $\ket{F=1, m_F}$ and $\ket{F=2, m_F}$, and develop the ``6-levels model'' depicted in Fig.~\ref{fig:link}. We believe this model captures the relevant physics of superlattices dressed with one Raman coupling 
and  includes spurious couplings between main and main or secondary and secondary sites.
The coupling between levels trapped at different positions, i.e. belonging to different bands of the lattice, is possible only because we are transferring momentum via the lasers. 
Since we are working at fixed $L=0$ and $m_F$, we restrict the previous notation $\ket{0 \, \alpha \, k}$ to the shorter $\ket{F,k}$, the two quantum numbers being
the hyperfine manifold $F=1,2$ and the position where the center-of-mass wavefunction is trapped  (for the meaning of $k=1,2,3$ see Fig.~\ref{fig:link}).
The model is characterized by only three relevant $S_{k'k}$, as depicted in Fig.~\ref{fig:link}; couplings between neighbouring main sites are negligible. 
The Hamiltonian reads as follows:
\begin{eqnarray}
H = & & d \; \proj{1,2}{1,2} + \Delta \left( \proj{2,1}{2,1} + \proj{2,3}{2,3} \right) + \nonumber\\ 
 & + & (\Delta + d) \proj{2,2}{2,2} + \nonumber \\
& + & \Omega e^{- i \omega t} \left[ \ \
 S_{1,2} \left( \proj{2,2}{1,1} + \proj{2,2}{1,3} \right) \right. + \nonumber \\
& & \phantom{\Omega e^{- i \omega t}} + 
S_{1,1} \, \left( \proj{2,1}{1,1} + \proj{2,3}{1,3} \right) + \nonumber \\
&   &  \phantom{\Omega e^{- i \omega t}} +
S_{1,2}^* \left( \proj{2,1}{1,2} + \proj{2,3}{1,2} \right) + \nonumber \\
&   & \phantom{\Omega e^{- i \omega t}} +
 \left.  S_{2,2} \,  \proj{2,2}{1,2}  \ \right]  \; + \; \text{H.c.}
\label{eq:6levelsmodel}
\end{eqnarray}
Once we apply the unitary transformation $\Gamma(t) = \exp [i \, d \left( |1,2\rangle \langle 1,2 | + |2,2\rangle \langle 2,2 | \right) t]$, the three levels $\ket{1,k}$ 
become degenerate.
In case the three inequalities $|S_{i,j} \Omega| / (\delta - d) \ll 1$ are fulfilled,
it is possible to use second-order perturbation theory in order to develop an effective Hamiltonian describing the dynamics within this submanifold:
\begin{eqnarray}
H_{pert} / \Omega^2 & = & - \left( \frac{|S_{1,1}|^2}{\delta-d} + \frac{|S_{1,2}|^2}{\delta} \right) [\proj{1,1}{1,1} + \proj{1,3}{1,3}] \nonumber \\
&  & - \left( \frac{|S_{2,2}|^2}{\delta - d} + 2 \frac{|S_{1,2}|^2}{\delta-2d}\right) \proj{1,2}{1,2}  \nonumber \\
&  & - \frac{|S_{1,2}|^2}{\delta} \proj{1,3}{1,1} \;+\; \text{H.c.} \nonumber \\
&  & - \left[ \frac{ S_{1,2} \, S_{1,1}}{2} \left( \frac{1}{\delta - d} + \frac{1}{ \delta-2d} \right) e ^{i d t} \, + \right. \nonumber \\
&  & + \left. \frac{ S_{2,2}^* \, S_{1,2}}{2} \left( \frac{1}{\delta - d} + \frac{1}{ \delta} \right) e ^{i d t}  \right] \cdot \nonumber \\
& & \cdot \, [\proj{1,2}{1,1} + \proj{1,2}{1,3} ]   \;+\; \text{H.c.}
\end{eqnarray}
Using this Hamiltonian we study the transfer rate of population from level $\ket{1,1}$ and $\ket{1,3}$ and viceversa. The main contribution is the direct coupling 
\begin{equation}
J^{(1)}_{13} = - \frac{|S_{1,2}|^2 \Omega^2 }{ \delta}. 
\label{eq:hoprate1}
\end{equation}
A second contribution, which in our system will prove to be not-negligible, comes from a sort of ``adiabatic elimination'' of the level $\ket{1,2}$:
\begin{equation}
J^{(2)}_{13} = - \frac{ \bra{1,3}H_{pert}\ket{1,2} \, \bra{1,2} H_{pert} \ket{1,1}}
{\bra{1,2} H_{pert} \ket{1,2} - \bra{1,1} H_{pert} \ket{1,1} + d} .
\label{eq:hoprate2}
\end{equation}

\begin{figure}[t]
\begin{center}
\includegraphics[width=0.9\columnwidth]{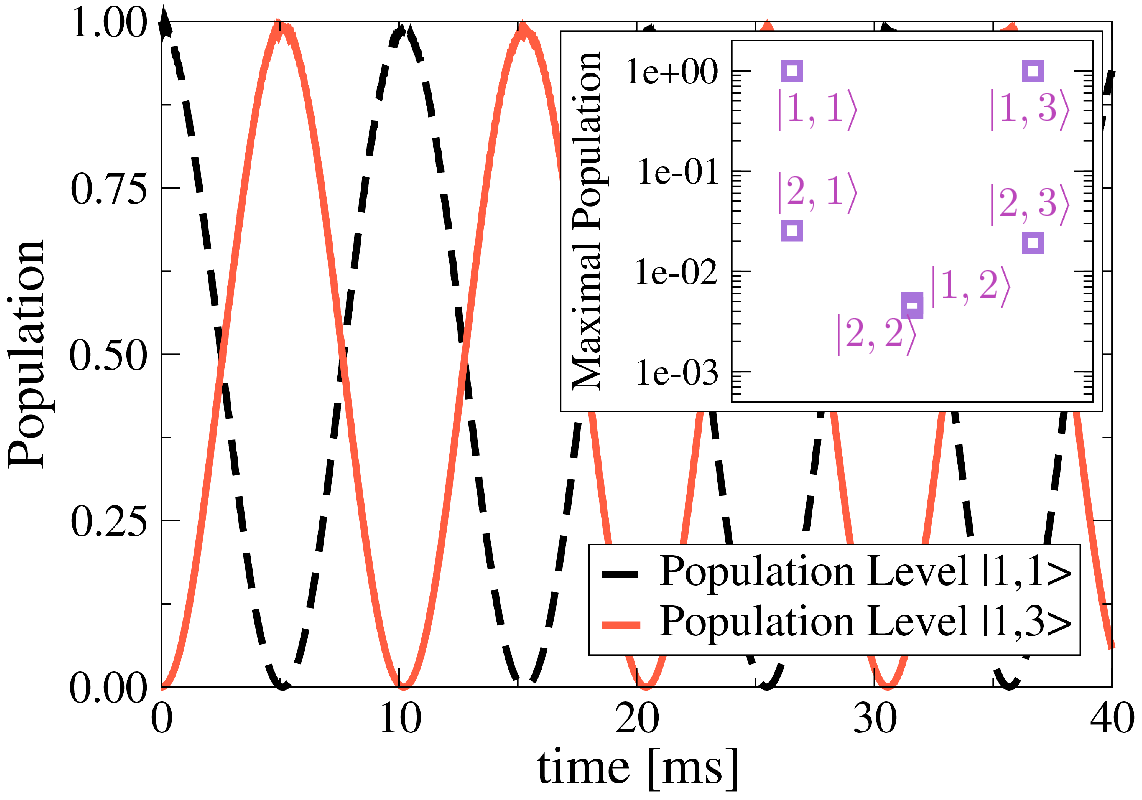}
\caption{Exact time evolution of the populations with $F_z=0$ of  the $6\times3$-levels model describing the dynamics of the hyperfine levels of the ground state under the action of three Raman couplings. The initial state is $\ket{1,1}$. Inset contains the maximum population reached in each level.
The parameters used are listed in Table~\ref{table:3times6model}.}
\label{fig:timeevol_6lev}
\end{center}
\end{figure} 

The main approximations in the presented ``6-levels model'' reside in the neglection of delocalized higher-energy free states and in the neglection of couplings between neighbouring higher-bands localised wannier functions. These processes could both induce spurious population transfers to next-neighbouring states. Regarding the first issue, this mainly means that experimentally there is a trade-off between a large detuning regime, allowing powerful lasers and strong effective couplings with noisy spurious population transfers, and a small detuning one,  with small clean couplings. Concerning the second point, this essentially implies a trade-off between a deep lattice configuration with localized wavefunctions  and a shallow lattice regime, with spread ones. In the former case neighbouring wannier functions are not connected by the Hamiltonian, but the overlap between different bands $S_{1,2}$ is also negligible;  in latter case the $S_{1,2}$ becomes important but couplings between neighbours become also significative.

It is possible to engineer the previous setup in order to get an independent control on the hopping rates of the different spin species. 
To this aim, we must be able to independently couple desired pairs of states (physical $F=1$ and auxiliary $F=2$ states) via independent Raman transitions. This can be realized with the help of  ``energy selection rules'', i.e. choosing pairs of states with distinct energy differences and
with the help of Raman transitions far-detuned from all the energy differences but that of the pair that they should couple.
In our case we take advantage of the fact that the hyperfine Land\'e factors of the manifolds $F=1$ and $F=2$ are one the opposite of the other, and 
suggest to split the hyperfine manifolds with a magnetic field and to use the level $\ket{\alpha = (F=2, F_z=m)}$ as ancilla state for the $\ket{\alpha = (1, m)}$ level. The energy differences of these pairs can then be detuned of circa $100$ MHz with moderate fields of circa $66$ G; in this case the use of Raman couplings detuned from the pair transition of  hundreds of kHz or even MHz would do the job (see Fig.~\ref{fig:splittings}).

\begin{table}[b]
\begin{ruledtabular}
\begin{tabular}{|c|l|| cc|}
Level \& $F_z$ & Energy & Parameters & \\
\hline 
$\ket{1,1}$, $m$ & $ \mu_F B m$ & $\Delta_{HF}$ & $6.8$ GHz \\
$\ket{1,2}$, $m$ & $ \mu_F B m + \delta$ & $\mu_F B$ & $50$ MHz \\
$\ket{1,3}$, $m$ & $ \mu_F B m$ & $\delta$ & $60$ kHz\\
$\ket{2,1}$, $m$ & $\Delta_{HF} - \mu_F B m$ & $S_{1,2}$ & $0.2$\\
$\ket{2,2}$, $m$ & $\Delta_{HF}- \mu_F B m + \delta$ & $S_{1,1}$ & $0.6$\\
$\ket{2,3}$, $m$ & $\Delta_{HF}- \mu_F B m$ & $S_{2,2}$ & $0.6$\\
\end{tabular}
\vspace{0.1cm}
\begin{tabular}{|c|c|c|c|}
\# Raman & $\Omega$ & $\omega$ & ang.mom. \\
\hline
1 & $31$ kHz  & $\Delta_{HF} - 2 \mu_F B + \delta - 300$ kHz  & 0\\
2 & $45$ kHz & $\Delta_{HF}  \; \phantom{- 2 \mu_F B} \; + \delta - 300$ kHz  & 0\\
3 & $39$ kHz & $\Delta_{HF} + 2 \mu_F B + \delta - 300$ kHz  & 0\\
\end{tabular}
\vspace{0.1cm}
\begin{tabular}{|c|cc|cc|}
$F_z$ & $J^{(1)}_{13} $ & $J^{(2)}_{13} $ & Extimated T & Numerical T \\
\hline
-1 & $-135.0$ Hz & $-18.3$ Hz & $20.4$ ms & $22.1$ ms\\
0 & $-270.0$ Hz & $-73.0$ Hz & $9.1$ ms & $10.1$ ms\\
+1 & $-202.5$ Hz & $-41.4$ Hz & $ 12.8$ ms & $14.0$ ms\\
\end{tabular}

\end{ruledtabular}
\caption{Parameters used in the simulation of the $3\times6$-levels model. We compare the periods of the Rabi oscillations with the theoretical values calculated taking into account only the Raman coupling which is supposed to drive the transition. Discrepancies are contributions of the off-resonant Raman couplings and higher order corrections.
}
\label{table:3times6model}
\end{table}

\begin{figure}[t]
\begin{center}
\includegraphics[width=0.9\columnwidth]{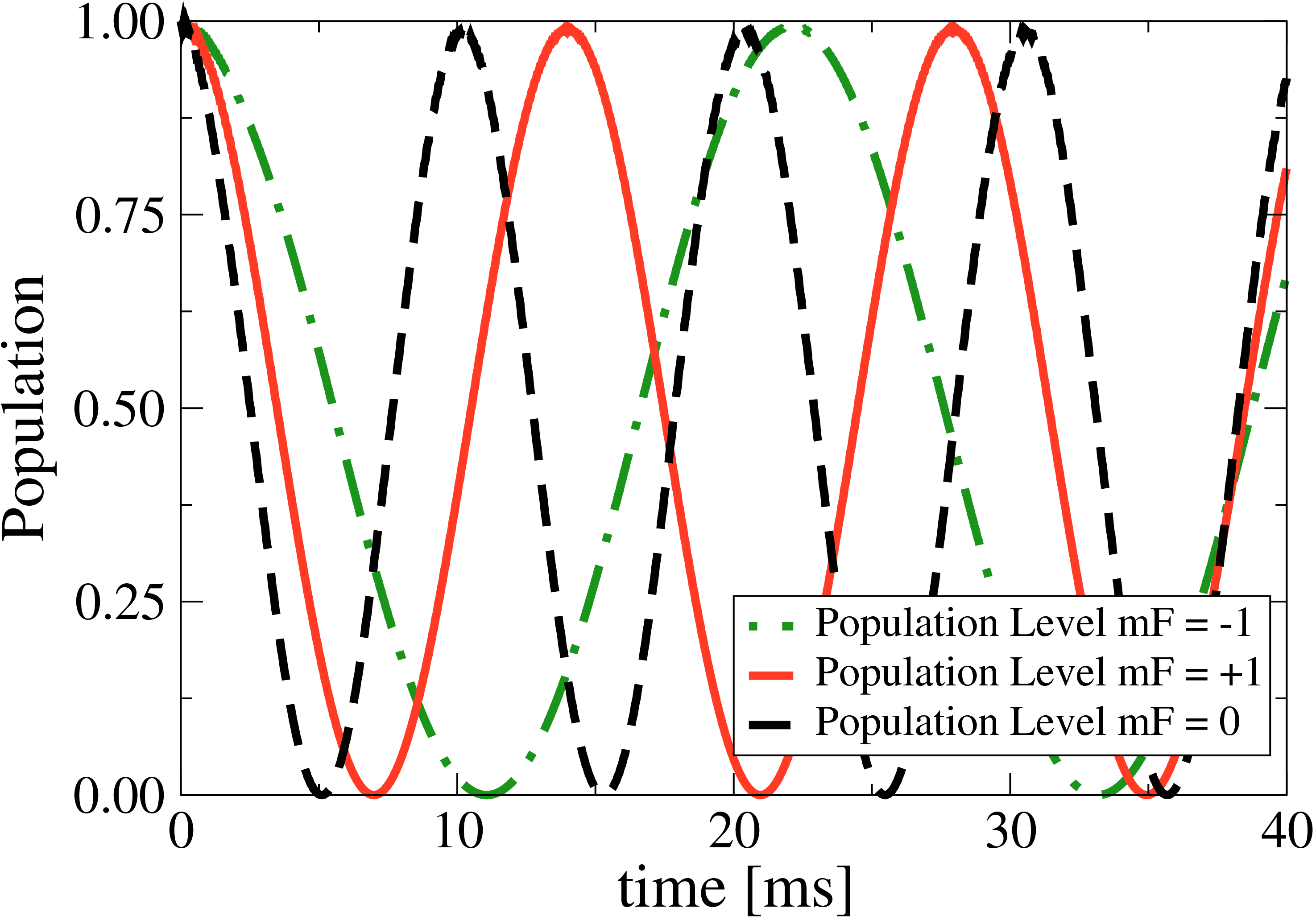}
\caption{Time evolution of the $\ket{1,1}$ levels with different magnetic numbers. As explained in the text,
with the parameters listed in Table~\ref{table:3times6model}
it is possible to tune  the different hopping rates to very different values.}
\label{fig:evolution_compare}
\end{center}
\end{figure}

We now report on some numerical simulations which consolidate the arguments given above and show that it is indeed possible to induce different hopping rates for the three spin species. We calculate the exact time evolution of a $6 \times 3$-levels model under the action of three different Raman couplings. The parameters characterizing the levels and the couplings are listed in Table~\ref{table:3times6model}.
We assume the possibility of engineering Raman couplings carrying no angular momentum ($\pi$ transitions) and therefore neglect the possibility of transferring population between states with different $F_z$. This factorizes our $18$-levels problem into three $6$-levels problems, which are numerically less demanding. In case this were not possible, spin mixing would still be almost prevented by energy-conservation constraints.

We show in Fig.~\ref{fig:timeevol_6lev} the exact time evolution of the six levels with $F_z=0$. At the beginning all the population is in the level $\ket{1,1}$ and very clear Rabi oscillations between the $\ket{1,1}$ and $\ket{1,3}$ levels can be seen. It is very important to notice that only a negligible fraction of the population is lost into the other four states (see the inset in Fig.~\ref{fig:timeevol_6lev}).

In Fig.~\ref{fig:evolution_compare} we compare the time evolution of the levels $\ket{F_z=-1,0,+1, k=1}$: it is very interesting to observe that  it is possible to induce different hopping rates for the three spin species. Indeed, as shown in Table~\ref{table:3times6model}, the simple application of Eq.~\ref{eq:hoprate1} and \ref{eq:hoprate2} corroborates the intuition that each hopping rate is ruled by only one Raman coupling, whereas the action of the others, far detuned, introduces only small corrections. It is also possible to check that including the other lasers the agreement with the experimental data improves.

Concluding, this analysis shows that optical lattices loaded with alkaline atoms display a hyerarchy of energy scales which could be exploited to engineer hopping operators breaking the SU(2) symmetry. In particular, we propose to employ the upper hyperfine manifold to provide auxialiary states to be adiabatically eliminated and to take advantage of superlattice configurations to trap them in the middle of each link.

\section{Magnetic Flux Quantization Condition \label{app:Flux}}

In Sec.~\ref{sec:pfaffian} 
we deal with a (discretely) translational invariant two dimensional lattice pierced by an external (homogeneous) magnetic field.
In this Appendix we provide more details on the study of such a system via a finite lattice with periodic boundary conditions (PBC).
In particular, we show that the need for mutual commuting Hamiltonian
and discrete-translation operator imposes non-trivial conditions on the dimension of the sample.

\subsection{Bulk}

We start discussing the Hamiltonian and the discrete translation operator in the bulk. We consider the Landau gauge:
$\mathbf A = B (0,x)$ and introduce the number of fluxes crossing each plaquette $\alpha = B a_x a_y / \Phi_0$ where $\Phi_0$ is the flux quantum and $a_x$ and $a_y$ are the dimensional lattice constants. From now on $x$ and $y$ will just be adimensional integer numbers labelling the sites of the lattice.

The standard generalization of the Bose-Hubbard Hamiltonian in presence of an external magnetic field is the Harper Hamiltonian:
\begin{eqnarray}
H = -J\sum_{x,y} \left[ e^{  - 2 \pi i \alpha x } d_{x, y+1}^{\dagger} d_{x,y}  + d_{x+1, y}^{\dagger} d_{x,y} \right] + H.c.
\phantom{ci}
\label{eq:hamiltonianconfasi}
\end{eqnarray}
where $d_{x,y}$ and $d_{x,y}^{\dagger}$ are boson annihilation and creation operators satisfying $[d_{x,y}, d_{x',y'}^{\dagger}] = \delta_{x x'} \delta_{y y'}$.

The action of the standard discrete translation operator $T_{m,n} = T(m a_x + n a_y)$ ($m,n \in \mathbb N$) on the field operators is the following:
\begin{eqnarray}
T_{1,0} \; d_{x,y}^{(\dagger)} \; T_{1,0}^{\dagger} \; &=& \; d_{x+1,y}^{(\dagger)} \nonumber \\
T_{0,1} \; d_{x,y}^{(\dagger)} \; T_{0,1}^{\dagger} \; &=& \; d_{x,y+1}^{(\dagger)} \nonumber \\
T_{m,n} \; &=& \;  T_{1,0}^m  \;  T_{0,1}^n = T_{0,1}^n \; T_{1,0}^m    
\nonumber
\end{eqnarray}

Since these $T_{m,n}$ operators do not commute with the Hamiltonian in Eq.~\ref{eq:hamiltonianconfasi}, we need a ``magnetic'' translation operator $M_{m,n}$ commuting with the Hamiltonian, which are the discrete version of the continuum case discussed in Ref.~\cite{PhysRev.134.A1607}:
\begin{eqnarray}
M_{1,0} \; d_{x,y}^{(\dagger)} \; M_{1,0}^{\dagger}  &=&  e^{+ (-) 2 \pi i \alpha y } \;  d_{x+1,y}^{(\dagger)} \nonumber \\
M_{0,1} \; d_{x,y}^{(\dagger)} \; M_{0,1}^{\dagger}  &=&  d_{x,y+1}^{(\dagger)} \nonumber \\
M_{m,n} d^{(\dagger)}_{x,y} M_{m,n}^{\dagger} &=& e^{- (+)i \pi \alpha m n} \; M_{1,0}^m \;   M_{0,1}^n
d_{x,y}^{(\dagger)} M_{0,1}^{\dagger n} \; M_{1,0}^{\dagger m}\nonumber \\
&=& e^{+ (-)i \pi \alpha m n}  \;   M_{0,1}^n \; M_{1,0}^m
d_{x,y}^{(\dagger)}  M_{1,0}^{\dagger m} \; M_{0,1}^{\dagger n} \nonumber
\end{eqnarray}
Last equation indicates clearly the peculiarity of the magnetic translations which leads to the Aharonov-Bohm effect: the result of a translation from one point to another one strongly depends on the followed path and eventually, moving along a closed loop, gives to the state a phase proportional to the encircled magnetic flux.

We can verify the commutativity of the magnetic translation operator with the Hamiltonian just by checking $M_{1,0}$ because translations along other directions commute straightforwardly:\\
\begin{eqnarray}
& & M_{1,0} \; e^{- 2 \pi i \alpha x } d_{x, y+1}^{\dagger} d_{x,y} \; M_{1,0}^{\dagger} =  \nonumber \\
& & \phantom{ciao ciao ci} = e^{- 2 \pi i \alpha x } e^{- 2 \pi i \alpha (y+1)} e^{+ 2 \pi i \alpha y}  d_{x+1, y+1}^{\dagger} d_{x+1,y}  \  \nonumber \\
& & \phantom{ciao ciao ci} = e^{- 2 \pi i \alpha (x+1)}  d_{x+1, y+1}^{\dagger} d_{x+1,y}   \nonumber \\
& &  M_{1,0} \; d_{x+1, y}^{\dagger} d_{x,y} \; M_{1,0}^{\dagger} = d_{x+1, y}^{\dagger} d_{x,y} \nonumber \\
& &  \Rightarrow M_{1,0} \; H \; M_{1,0}^{\dagger} = H \nonumber
\end{eqnarray}
where we exploit the sum over dummy $x$ in $H$ and change variables to $x' = x+1$, always possible in the bulk.

\subsection{Boundaries}

We now discuss the possibility of studying the previous infinite system with a finite system
of dimension $L_x \times L_y$ with PBC. $L_{x,y}$ are here adimensional numbers which can be used to define the total number of fluxes crossing the finite system: $N_{\Phi} = L_x L_y \alpha$.
In order to be able to identify the bosonic operators residing on sites whose distance is $m L_x a_x + n L_y a_y$, with $m,n \in \mathbb N$, we must require the total number of fluxes $N_{\Phi}$ to be an integer number. This can be proven simply translating the field operator around one plaquette $L_x \times L_y$.
As before, we also require the Hamiltonian and the ``magnetic'' translation operators to commute; in particular, we discuss in detail the interesting case of translation along $\hat x$: $ M_{1,0} \; H \; M_{1,0}^{\dagger} \; = \; H$.

We separately analize 
this equation on each link of the finite lattice. In particular, when considering the   links  oriented along the $\hat y$ direction, it reduces to the following equality:
\begin{eqnarray}
e^{- 2 \pi i \alpha \, x} \; e^{-2 \pi i \alpha \, [\overline{y+1} - y]} \;
\cdot   d^{\dagger}_{\overline{x+1}, \overline{y+1}}
d_{\overline{x+1}, y} = \nonumber \\
=e^{- 2 \pi i \alpha \, \overline{x+1}} \; d^{\dagger}_{\overline{x+1}, \overline{y+1} } d_{\overline{x+1}, y}
\label{eq:dafulfillare}
\end{eqnarray}
where $\overline{x+1}$ denotes the modulus count $\left( x+1 \mod L_x \right)$; the same holds for $y$.

We distinguish four cases:
\begin{enumerate}
\item $ x \in [0, L_x-2] \;  \wedge \; y \in [0, L_y -2]$: Eq.~\ref{eq:dafulfillare} is automathically satistified, as it happens in the bulk;
\item $x = L_x -1 \; \wedge \; y \in [0, L_y -2]$: Eq.~\ref{eq:dafulfillare} is fulfilled only if $e^{-2 \pi i \alpha (L_x -1)} e^{ - 2 \pi i \alpha} = 1$, which implies $\alpha L_x \in \mathbb N$;
\item $x \in [0, L_x-2] \; \wedge \; y =L_y - 1$: Eq.~\ref{eq:dafulfillare} is fulfilled only if 
$ e^{+2 \pi i \alpha (L_y -1)}  = e^{-2 \pi i \alpha }$, which implies $\alpha L_y \in \mathbb N$;
\item $x = L_x-1 \; \wedge \; y = L_y - 1$: Eq.~\ref{eq:dafulfillare} is fulfilled only if $e^{-2 \pi i \alpha (L_x -1)} e^{+ 2 \pi i \alpha (L_y -1 )} = 1$, which implies $\alpha (L_y - L_x )\in \mathbb N$. 
\end{enumerate}

The double constraint $N_\Phi / L_x, \; N_\Phi / L_y \in \mathbb{N}$ and the desired magnetic filling one $N_\Phi = N$ strongly
reduces the number and variety of finite size systems numerically treatable with moderate effort.
The Hilbert space for the examined $4 \times 4$ lattice with $4$ particles consists of $3.620$ states, but already $5$ three-hardcore bosons
on a $5 \times 5$ grid need $110.630$ to be described. 
Such strict constraints could be circumvented if one introduces proper singularities of the magnetic field
to fulfill the correct translational and periodic conditions;
however, any spurious correction introduced by hand would strongly affect the numerics on the small scales treatable.
Therefore we decided to stay stuck to the strictest version given above.
An extensive numerical study of this problem, though interesting, goes well beyond the purposes of the present paper and is left for future investigations.

\bibliography{3body.bib}
\end{document}